\documentclass[a4paper]{article}

\usepackage[english]{babel}
\usepackage[utf8]{inputenc}
\usepackage[round]{natbib}
\usepackage{authblk,bbm,bm}
\usepackage{amsmath,setspace,geometry,lineno,amssymb}
\usepackage{graphicx,url}
\usepackage[colorinlistoftodos]{todonotes}
\usepackage{algorithmic}
\usepackage[boxed]{algorithm2e}
\usepackage{listings}
\newcommand{\normal}{\text{N}}

\begin{document}
\begin{spacing}{1.15}

\title{Statistical modelling of individual animal movement: an overview of key methods and a discussion of practical challenges}
\author{
  Toby A.\ Patterson\\
  \vspace{-1em}
  \textit{CSIRO Oceans and Atmosphere, Hobart, Australia} \\
  \and
  \vspace{-1em}
  Alison Parton\\
  \vspace{-1em}
  \textit{School of Mathematics and Statistics, University of Sheffield, UK}\\
  \and  
    \vspace{-1em}
  Roland Langrock\\
  \vspace{-1em}
  \textit{Department of Business Administration and Economics, Bielefeld University, Germany} \\
  \and
  \vspace{-1em}
  Paul G.\ Blackwell\\
  \vspace{-1em}
  \textit{School of Mathematics and Statistics, University of Sheffield, UK} \\
\and
  \vspace{-1em}
    Len Thomas\\
  \vspace{-1em}
  \textit{School of Mathematics and Statistics, University of St Andrews, UK} \\
  \and
  \vspace{-1em}
  Ruth King\\
  \vspace{-1em}
  \textit{School of Mathematics, University of Edinburgh, UK}  \\
 }

\date{}
\maketitle

\begin{abstract} 
With the influx of complex and detailed tracking data gathered from electronic tracking devices, the analysis of animal movement data has recently emerged as a cottage industry amongst biostatisticians. New approaches of ever greater complexity are continue to be added to the literature. In this paper, we review what we believe to be some of the most popular and most useful classes of statistical models used to analyze individual animal movement data. Specifically we consider discrete-time hidden Markov models, more general state-space models and diffusion processes. We argue that these models should be core components in the toolbox for quantitative researchers working on stochastic modelling of individual animal movement. The paper concludes by offering some general observations on the direction of statistical analysis of animal movement. There is a trend in movement ecology toward what are arguably overly-complex modelling approaches which are inaccessible to ecologists, unwieldy with large data sets or not based in mainstream statistical practice. Additionally, some analysis methods developed within the ecological community ignore fundamental properties of movement data, potentially leading to misleading conclusions about animal movement. Corresponding approaches, e.g.\ based on L\'evy walk-type models, continue to be popular despite having been largely discredited. We contend that there is a need for an appropriate balance between the extremes of either being overly complex or being overly simplistic, whereby the discipline relies on models of intermediate complexity that are usable by general ecologists, but grounded in well-developed statistical practice and efficient to fit to large data sets. 
\end{abstract}
\vspace{0.5em}

\vspace{0.5em}

\section{Introduction} 
Movement ecology seeks to infer why organisms move through space and what constraints operate on them as they do. It is the study of how movement shapes their overall ecology, and the factors, both intrinsic and extrinsic, that influence movement.  Movement ecology broadly concerns the study of both animals and plant movement and dispersal. While there are commonalities in some aspects of their study, there are also many differences. In this paper we concern ourselves purely with individual organismal movement as measured through instruments and sensors which record position at relatively short time scales.  Although it is possible that some of the aspects we discuss also apply to plants, practically this means we are nearly always referring to the study of animal movement. 


The discipline has been driven by the possibility of telemetering the paths of free-moving animals navigating their natural habitats. Devices such as satellite and GPS tags, radio tracking, radar and acoustic monitoring have generated large and, to ecologists and statisticians alike, largely unfamiliar data sets. Each technology has its limitations and strengths in terms of accuracy, frequency and longevity. One fundamental feature of movement ecology is that it takes a `bottom-up' approach to understanding population processes: it works by tracking individuals and seeking to infer properties of populations. This brings several challenges, not least that these individual-level data do not sit easily within the remit of the conventional biometric toolbox employed by field ecologists. Statistically, movement processes are reasonably described as noisy, non-linear and highly spatially and temporally correlated. As a result, ecologists and statisticians alike have searched for new tools for understanding animal movement. 

Yet, movement ecology has also grappled with its conceptual foundations. Despite being widely recognized as a fundamental process governing population dynamics, the rationale for studying movement can be somewhat ill-defined. In other areas of ecological data analysis, say estimation of animal abundance, the motivation is clear, namely to determine population size. Moreover, the reasons for wanting such an estimate are clear. We maintain that this is not always the case in movement ecology. While observations of movement, especially for a new species, are generally interesting, the use of movement analyses, especially in applied ecological settings such as conservation or management, is not always coherently stated. Additionally, movement studies often suffer from low sample sizes and, even in an informal sense, a lack of statistical design \citep{patterson2011designing,mcgowan2016integrating}. While these problems can be difficult or even impossible to overcome, this is of obvious importance for arriving at a set of statistical methods which are suitable for purpose.  


Statistical methods for the analysis of individual movements can be problematic in at least two ways. First, we contest that some of the models being developed arguably are overly complex, both structurally and in terms of the machinery required to fit them. In these cases, the complexity can be beyond what is required to address the study goals. Care is therefore needed that complex models are not constructed based on data from a small number of individuals or of short duration. The danger of this is that much effort is expended on capturing aspects of a possibly unrepresentative data set. At the other extreme lies another problem, namely that some movement models are hopelessly simplistic, e.g.\ relying on a single-parameter model to describe a vast array of complex behaviours. Yet, literal interpretation of the mathematical properties of such simple models have been offered as evidence for strong claims about animal movement. 

While similar issues could be identified in many scientific endeavours, there is a need for movement ecology to recognise these problems. Addressing these requires the discipline clarifying its collective aims and consciously seeking to build workable and well-understood analysis approaches that can be widely applied. Part of this process is the identification of a set of models (a) that are reasonably appropriate to the nature of most individuals' movement data and associated research questions, (b) whose statistical properties are well understood and (c) that are sufficiently computationally efficient that they may be applied to representative, i.e.\ sufficiently large, data sets. 

Real animal movements and behaviour are of course highly complex and dynamic. There is a limit to what can be observed from position and sensor data alone. Therefore, to parse out gross features from the data, it often makes sense to assume movement processes to be driven by switches between behavioural modes, and several of the modelling approaches that we discuss below do allow for different phases or modes of movement. There is a mounting number of papers which seek to make interpretation of animals' movements tractable by assuming that they typically move in a set of movement modes---e.g.\ rapid movements between regions (``transit'' or ``exploratory'' movements) vs.\ highly resident (``encamped'') movements which are related to activities such as resting or foraging. These approaches are at the core of what we cover here. 

From the outset we admit that our treatment is myopic in the sense that several other widely used statistical approaches are not discussed in detail---in particular those that look at broader spatial and temporal scales, at which behavioural state switching is most likely irrelevant.  If considered at all, we provide only a brief overview of those other approaches, and instead focus on three types of models---hidden Markov models (HMMs), more general state-space models (SSMs) and diffusion processes--- in much more detail. We identified these as key tools for conducting statistical analyses of animal movement data collected at the individual level. Accordingly the paper is therefore restricted mostly to consideration of models that analyze trajectories (or metrics derived from them). Most commonly these trajectories are expressed as time series of geographical coordinates. Our lack of attention to other areas of movement ecology, or to ecological settings where movement is important, such as resource and habitat selection, is simply because we regard these as related, but separate branches of the discipline, characterized by different statistical problems and associated techniques. We also note that even when we consider animal trajectory data alone, many telemetry devices record concurrent sensor data that are useful for extracting behavioural signals. For instance, most instruments deployed on air-breathing marine predators (marine mammals, seabirds) collect not only position estimates but also data describing diving behaviour. These sorts of data are obviously relevant to characterizing the state of the animal, and in our view are often not considered in sufficient detail in SSMs (to name an example). In acknowledging this issue, we must also admit that we do not, in this review paper, consider in any detail techniques or models to tackle these combined data, although we note in passing that combining types of data is perhaps less of a conceptual and technical jump from existing techniques than is often appreciated. For an example of an analysis of such a more complex data set using essentially only standard methods, see \citet{DeRuiter2016}, where time series comprising seven data streams, corresponding to different measures of blue whale activity, are analyzed using a joint state-switching model.

The goal of this paper is therefore to provide a relatively detailed examination of selected methods for analysis of individual animal tracking data. Our intended audience are the statistically-minded ecologists and ecologically-minded statisticians who are actively working with this data, day to day, although we hope that the material below might also provide a technically honest entry point for the uninitiated. This is in contrast to previous reviews in this area \citep[e.g.][]{Patterson2009}, which have been for a broader ecological audience and by necessity have needed to omit the technical intricacies. 

\section{Animal movement data}


Data sets on animal movement typically contain positions in space over a sequence of discrete points in time, observed by using Global Position System (GPS) telemetry technology, for instance. For land-based animals, this will usually be in the horizontal plane, while for aerial and marine life, geographical space studies also exist in one dimension, such as vertical movements of aquatic species---this choice clearly being dependent on the questions being raised. A few studies exist of where high resolution three-dimensional movement data is available \citep[e.g.][]{laplanche2015tracking}, but these are less common and we do not consider them in this review. Sampling of locations is often made at regular time intervals, and the hidden Markov model (HMM) and state-space model (SSM) approaches described below are, with some important exceptions, restricted to such data and are not generally well suited to observations that are irregularly spaced in time (but see Section \ref{BayesSSM}). In contrast, continuous-time approaches, such as those based on diffusion models (Section \ref{DM}), straightforwardly accommodate irregular time intervals, whether they arise by design, through missing data, or through the limitations of the sensor technology.

Sampling intervals vary considerably across studies, ranging from fractions of seconds up to days. The time difference between observations affects what types of inference can be made and what modelling approaches can be applied. It is therefore important that care should be taken when choosing the sampling interval and that researchers think ahead to the necessary analysis prior to deployment of telemetry instruments. If the goal of an analysis is to infer the behavioural states of an animal, or proxies thereof, then observations need to be made at a temporal scale that is meaningful with regard to the behavioural dynamics of the animal. 

Various different movement metrics can be considered when modelling telemetry data. These include, but are not limited to: 
\begin{itemize}
\item the bivariate positions themselves or increments of these (velocity or displacement) in either dimension;
\item distances between successively observed positions (usually referred to as the {\em step lengths});
\item compass directions (headings);
\item changes of direction between successive relocations (usually referred to as the {\em turning angles}).
\end{itemize}
We note that, conditional on the position and heading of an animal at the initial observation, the bivariate series of step lengths and turning angles completely determines the entire subsequent movement path, and hence all metrics listed above. Step lengths and turning angles are often considered together when analysing and interpreting movement data, in particular because they lead to intuitive interpretations \citep{mar88}. Note, however, that each turning angle depends on a sequence of three consecutive observations, and so the arbitrary timescale of the observations is even more influential. 

While this review primarily discusses the case of location data, there are many other types of animal movement data, such as location derived from light sensors, accelerometers, magnetometers, measurements of bearing, etc.\ that are used to derive position information. Additionally, some deployments of instruments return a mixture of these. Finally, we note that a wide range of technologies exist that provide information on fine-scale behaviours indeterminable from locational data alone. Readers are directed to \citet{Cooke2004}, \citet{Cooke2013}, \citet{Rutz2009}, \citet{Wilmers2015} and \citet{leosbarajas2016} for detailed reviews.

\section{Overview of individual-level models for animal movement}
\label{overview}

In most cases, the type of data at hand largely dictates which modelling approach to use. Given data, a decision whether to use HMMs, SSMs or diffusion processes can broadly be summarised as follows.

If the data are collected at regular sampling units, e.g.\ hourly, daily or every time a marine mammal comes to the surface to breathe, then most often a discrete-time model would be used. If in addition the measurement error is negligible, then HMMs represent a natural, accessible and most likely computationally feasible approach, which would typically be used to make inference for example on how animals interact with their environment (see Section \ref{HMMs}). If however the measurement error is non-negligible, then SSMs account for this, at the cost of an increase in complexity, regarding both the implementation and the computational effort. Like HMMs, SSMs can be used for making general inference, though in some cases they are applied simply to filter the noisy locations (see Section \ref{SSMs}).

If the data are not equally spaced, i.e.\ if there is no regularity in the sampling process, then continuous-time models such as diffusion processes constitute the most natural choice. Of course these models can also be applied to regularly sampled data. The main drawback of those models, from a user's perspective anyway, is that they are less accessible than HMMs and SSMs (see Section \ref{DM}). 

There are of course exceptions to the above crude classification of how different types of data are tied to specific modelling approaches. For example, for irregularly spaced data, instead of using a continuous-time approach, it has been suggested to interpolate the recorded locations on the required grid, then fitting an SSM that accounts for the corresponding error due to the interpolation. 


\section{Hidden Markov models: discrete time, no measurement error}
\label{HMMs}

\subsection{Model formulation}
\label{HMMformulation}

HMMs are natural candidates for modelling animal movement data. Indeed, HMMs have successfully been used to analyse the movement of, {\em inter alia}, caribou \citep{Franke2004}, fruit flies \citep{Holzmann2006}, tuna \citep{Patterson2009}, panthers \citep{vandeKerk2015}, woodpeckers \citep{McKellar2015} and white sharks \citep{towner2016}, to name but a few. 
One typically considers bivariate time series comprising step lengths and turning angles, {\em regularly spaced in time} and assumed to be {\em observed with no or only negligible error}. Within the HMM framework, such a time series is typically referred to as the {\em state-dependent process}, since each of the corresponding observations is assumed to be generated by one of $N$ distributions as determined by the state of an underlying hidden (i.e.\ unobserved) $N$-state Markov chain. The states of the Markov chain can be interpreted as providing rough classifications of the behavioural dynamics (e.g.\ more active vs.\ less active). 
In the following, we describe the key assumptions involved in basic HMMs and also how model fitting for this class of models can easily be accomplished. 
Almost all the methods described below are implemented in the recently released R package \texttt{moveHMM} \citep{michelot2015}.

\begin{figure}[!htb]
\vspace{1em}
\begin{center}
	\includegraphics[width=0.8\linewidth]{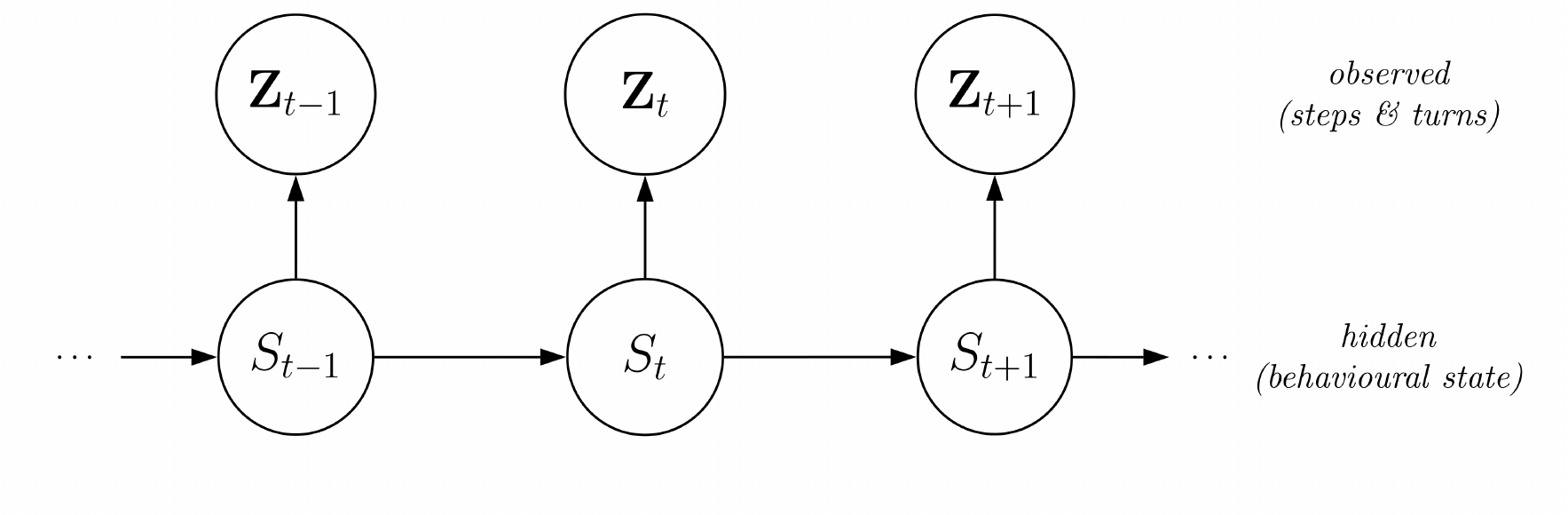}
	\caption{Modelling of bivariate time series comprising step lengths and turning angles using HMMs: illustration of the basic dependence structure as a directed acyclic graph.}
	\label{graph}
\end{center}
\end{figure}

Let the state-dependent process be denoted by $\{ \mathbf{Z}_t \}_{t=1}^T$, with realizations $\mathbf{z}_t=(l_t,\phi_t)$, where $l_{t}$ is the step length in the interval $[t,t+1]$ and $\phi_{t}$ is the turning angle between the directions of travel during the intervals $[t-1,t]$ and $[t,t+1]$, respectively.
Furthermore, let the underlying nonobservable $N$-state Markov chain be denoted by $\{ S_t \}_{t=1}^T$. In the most basic model, the dependence structure is such that, given the current state of $S_t$, $\mathbf{Z}_t$ is conditionally independent from previous and future observations and states, and the (homogeneous) Markov chain is of first order (Figure \ref{graph}). We summarize the probabilities of transitions between the different states in the $N \times N$ transition probability matrix (t.p.m.) $\boldsymbol{\Gamma}=\left( \gamma_{ij} \right)$, where $\gamma_{ij}=\Pr \bigl(S_{t}=j\vert S_{t-1}=i \bigr)$, $i,j=1,\ldots,N$.
The initial state probabilities are summarized in the row vector $\boldsymbol{\delta}$, where $\delta_{i} = \Pr (S_1=i)$, $i=1,\ldots,N$. 

HMMs for animal movement data typically involve the additional assumption of contemporaneous conditional independence \citep{Zucchini2009}. That is, conditional on the current state, the step-length and turning-angle random variables are assumed to be independent:
$$f_i (\mathbf{z}_t) = f(\mathbf{z}_t | S_{t}=i)= f \bigl( (l_t,\phi_t) | S_{t}=i\bigr)= f (l_{t} | S_{t}=i)f (\phi_{t} | S_{t}=i).$$ 
(Here and elsewhere we use $f$ as a general symbol for a density function.) This assumption greatly facilitates inference, yet is not overly restrictive. In particular, the two component series will still be mutually dependent, because the states induce dependence between the component series. 
Plausible distributions for modelling the circular-valued turning angles include the von Mises and wrapped Cauchy. Step lengths are, by nature, positive and continuous, which renders gamma and Weibull distributions plausible candidates for modelling this component. 

Most of the HMMs considered in movement ecology thus far involve a two-state Markov chain, where each state is associated with a different correlated random walk (CRW) pattern. CRWs involve correlation in directionality and can be expressed by a turning-angle distribution with mass centred either on zero (for positive correlation) or on $\pi$ (for negative correlation). The two states of the corresponding models are often associated with the animal being either `encamped' (with mostly short step lengths and many turnings) or `exploring' (with, on average, longer step lengths and more directed movement, as expressed by smaller turning angles). This kind of labelling of the states should be made with caution; it is generally accepted that these states merely provide convenient proxies of an animal's actual behavioural state (see Section \ref{states}). 

In movement ecology, it is usually of primary interest to relate the state-switching process to environmental covariates, thereby investigating how individuals interact with their environment. This can be achieved by considering a non-homogeneous Markov chain, with time-varying t.p.m.\ $\boldsymbol{\Gamma}^{(t)}=\left( \gamma_{ij}^{(t)} \right)$, linking the transition probabilities to the covariate vectors, $\mathbf{v}_{\cdot 1}, \ldots ,\mathbf{v}_{\cdot T}$, via the multinomial logit link: 
\begin{equation}\label{covtpm} \gamma_{ij}^{(t)}=\Pr \bigl(S_{t}=j\vert S_{t-1}=i \bigr)=\frac{\exp (\eta_{ij}) }{\sum_{k=1}^N\exp (\eta_{ik})},
\end{equation}
where $\eta_{ij} = \beta_{0}^{(ij)} + \sum_{l=1}^p \beta_{l}^{(ij)} v_{lt}$ if $i\neq j$ (and 0 otherwise). 


\subsection{Inference for HMMs}

For a homogeneous HMM, 
with parameter vector $\boldsymbol{\theta}$, the likelihood is given by
\begin{align}\label{bruteforceL}
\nonumber \mathcal{L}^{\text{HMM}}(\boldsymbol{\theta}) & = 
f(\mathbf{z}_1, \ldots, \mathbf{z}_T) \\
\nonumber & = 
\sum_{s_1=1}^N \ldots \sum_{s_T=1}^N f(\mathbf{z}_1, \ldots, \mathbf{z}_T | s_1, \ldots, s_T) f(s_1, \ldots, s_T) \\
& = \sum_{s_1=1}^N \ldots \sum_{s_T=1}^N \delta_{s_1} \prod_{t=1}^T f (l_{t} | s_{t})f (\phi_{t} | s_{t})  \prod_{t=2}^T \gamma_{s_{t-1},s_t} \, ,
\end{align}
where we exploited the dependence structure in the last step. In this form, the likelihood involves $N^T$ summands, rendering its evaluation infeasible even for a small number of states, $N$, and a moderate number of observations, $T$. This has led some users to believe that a Bayesian approach is required to fit an HMM to movement data, often leading to the use of WinBUGS. 
This approach, in which the data are augmented  with the states at each time point, is computationally inefficient because the Markov chains display high auto-correlation (see Section \ref{BayesSSM}). Consequently, fitting HMMs to large movement data sets is often infeasible when using Markov chain Monte Carlo (MCMC) in this way. 

Returning to the likelihood of an HMM, it turns out that the use of a recursive scheme called the {\em forward algorithm} leads to a much more efficient calculation than via brute force summation over all possible state sequences as in (\ref{bruteforceL}). To see this, we define the forward probability of state $j$ at time $t$ as $ {\alpha}_{t}(j) = f(\mathbf{z}_1,...,\mathbf{z}_{t},s_{t}=j)$, and the vector of forward probabilities at time $t$ as 
$\boldsymbol{\alpha}_{t} = \bigl( {\alpha}_{t}(1), \ldots, {\alpha}_{t}(N) \bigr)$. The key point now is that 
$\boldsymbol{\alpha}_{t}$ can be calculated based on $\boldsymbol{\alpha}_{t-1}$. More specifically, using the conditional independence assumptions it can easily be shown that
\begin{align*}
{\alpha}_{t}(j) 
& = \sum_{i=1}^N  {\alpha}_{t-1}(i) \gamma_{ij}  f_j(\mathbf{z}_t) .
\end{align*}
In matrix notation, this becomes $ \boldsymbol{\alpha}_{t} = \boldsymbol{\alpha}_{t-1} \boldsymbol{\Gamma} \mathbf{Q}(\mathbf{z}_{t})$,
where $\mathbf{Q}(\mathbf{z}_{t})= \text{diag} \bigl( f_1 (\mathbf{z}_{t}), \ldots, f_N (\mathbf{z}_{t}) \big)$. Together with the initial calculation
$ \boldsymbol{\alpha}_{1} = \boldsymbol{\delta} \mathbf{Q}(\mathbf{z}_{1})$, 
this is the forward algorithm. Rather than separately considering all possible hidden state sequences, as in (\ref{bruteforceL}), the forward algorithm exploits the dependence structure to perform the likelihood calculation recursively, traversing along the time series and updating the likelihood and state probabilities at every step. Such efficient recursive computation is one of the key reasons for the popularity and widespread use of HMMs -- the same trick can be applied to forecasting, state decoding (see later) and model checking.  The main price to pay for being able to rapidly conduct such analyses is that the Markov assumption needs to be made for the state process. This assumption is somewhat unrealistic in many applications, but is a good example of attempting to identify the correct balance between complex models that try to mimic as many aspects of the data as possible, often at the expense of becoming computationally intractable, and overly simplistic models such as the L\'evy walk which ignores almost all striking features of the data (see Section \ref{levy}).

The forward algorithm can be applied in order to first calculate $\boldsymbol{\alpha}_{1}$, then $\boldsymbol{\alpha}_{2}$, etc., until one arrives at $\boldsymbol{\alpha}_{T}$, the sum of all elements of which obviously yields the likelihood. Thus,
\begin{equation}\label{lik}
\mathcal{L}^{\text{HMM}}(\boldsymbol{\theta}) = \boldsymbol{\alpha}_{T}\mathbf{1}^t = 
\boldsymbol{\alpha}_{T-1} \boldsymbol{\Gamma} \mathbf{Q}(\mathbf{z}_{T}) \mathbf{1}^t = \ldots = 
\boldsymbol{\delta} \mathbf{Q}(\mathbf{z}_1) \boldsymbol{\Gamma} \mathbf{Q}(\mathbf{z}_2) \ldots \boldsymbol{\Gamma} \mathbf{Q}(\mathbf{z}_T) \mathbf{1}^t \, ,
\end{equation}
where $\mathbf{1}\in \mathbbm{R}^N$ is a row vector of ones. The computational cost of evaluating (\ref{lik}) is {\it linear} in the number of observations, $T$, such that a numerical maximization of the likelihood becomes feasible in most cases. 
Technical issues arising in the numerical maximization, such as parameter constraints and numerical underflow, are straightforward to deal with \citep{Zucchini2009}. 

A popular likelihood-based alternative is given by the expectation-maximization (EM) algorithm. The EM algorithm also involves an iterative scheme for finding the maximum likelihood estimate, by alternating between updating the conditional expectation of the states (given the data and the current model parameters), and updating the model parameters based on the complete-data log-likelihood where the unknown states are replaced by their conditional expectations. We do not elaborate on EM here, since we agree with \citet{mac14} in there being no apparent reasons to prefer it over direct likelihood maximization, which is easier to implement. 

From a Bayesian perspective, the efficient evaluation of the likelihood is also a great advantage. There is then no need to augment the data with the unknown states: we can simply use the likelihood as above, applying the forward algorithm, and carry out either a simple MCMC algorithm over the parameter space, such as random walk Metropolis-Hastings, or direct numerical maximisation over the (log) posterior density, if approximating the posterior distribution near its mode is adequate.

Of course, an analysis of movement data using HMMs does not end with estimating the model parameters, and the HMM toolbox offers a variety of additional inferential techniques. In particular, this includes the Viterbi algorithm, which is a recursive algorithm for ``state decoding'' --- i.e. identifying the most likely state sequence to have generated the observed time series, under the fitted model.
Furthermore, the forward and backward probabilities can be used to perform state prediction (i.e. calculate the state probabilities at a given time) and to calculate ``pseudo-residuals'' (also known as quantile residuals) for model checking. For more details, we refer to \citet{Zucchini2009}. 

\subsection{Real data example: daily movement of elk}

To illustrate the HMM approach to modelling animal movement, we re-analyze the elk data discussed in \citet{Morales2004}. 
The data set was downloaded from the Ecological Archives (\url{http://www.esapubs.org/archive/ecol/E085/072/elk_data.txt}).
We note that this new analysis is neither an attempt to replicate nor an attempt to improve the models discussed in detail by \citet{Morales2004}. The data set comprises four tracks, each with daily observations and several associated habitat covariates. There are $735$ observed locations in total. For more details, see \citet{Morales2004}. 

\begin{figure}[!htb]
\begin{center}
	\includegraphics[width=0.65\linewidth]{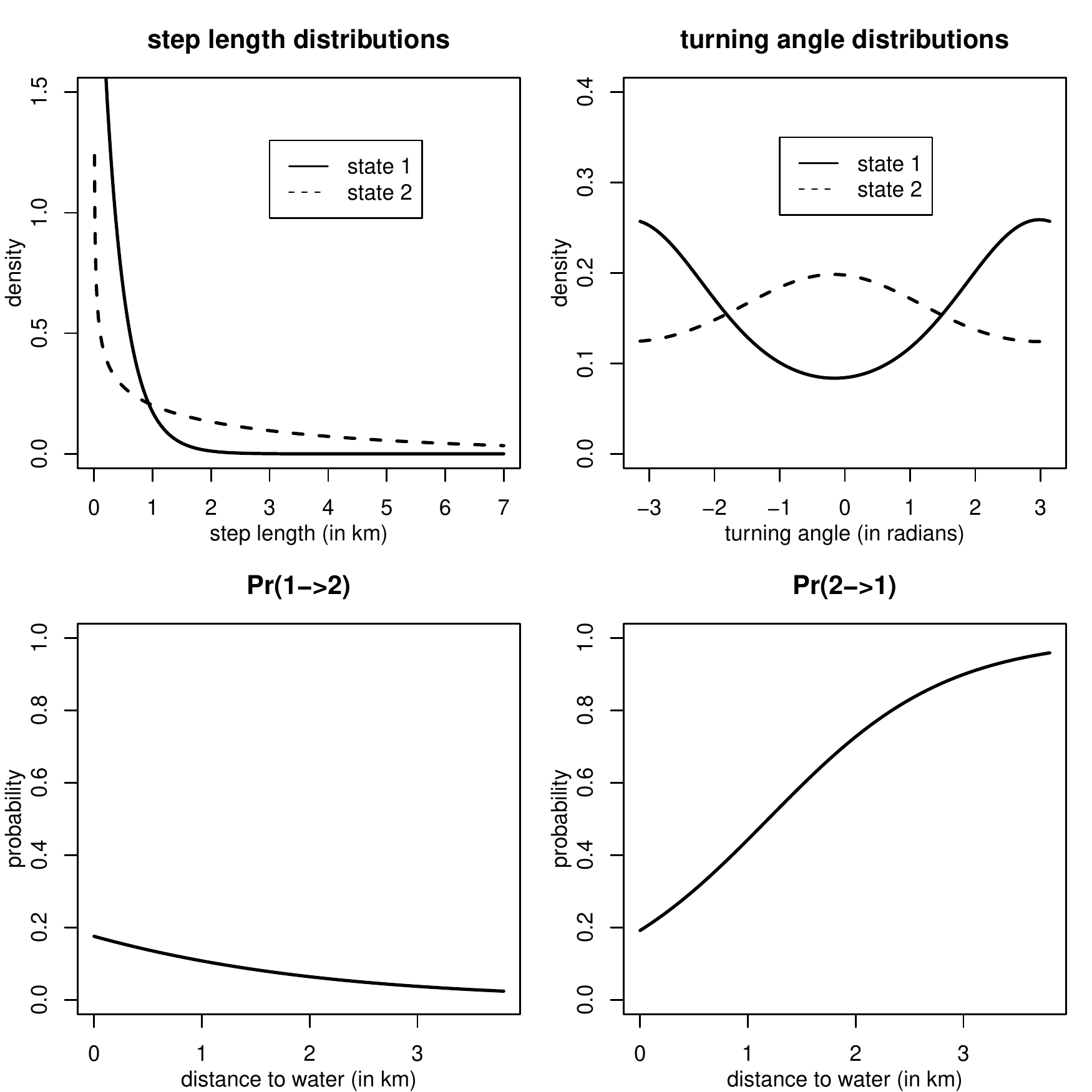}
	\caption{HMM applied to elk movement data: fitted state-dep.\ distributions (top row) and estimated effect of covariate `distance to water' on state-switching dynamics (bottom plot).}
	\label{elkfig}
\end{center}
\end{figure}
 
We fitted a joint two-state HMM, with von Mises turning angle and gamma step length distributions, to all four elk's tracking data, assuming all model parameters to be common to all individuals. 
About 2\% of the step lengths were exactly equal to 0, which we accounted for following  \citet{McKellar2015} by including additional parameters specifying state-dependent point masses on 0 in the otherwise strictly positive (gamma) step length distributions.
To illustrate the type of ecological inference that can be made using HMMs, we additionally implemented an AIC-based forward selection of covariates influencing the state-switching dynamics, as in (\ref{covtpm}), which led to the inclusion of exactly one covariate, namely ``distance to water'' ($\Delta$AIC compared to the baseline model without covariates: $11.3$). This latter type of inference, i.e.\ the fact that HMMs can easily be used to relate the evolution of an animal's behavioural states to environmental and habitat conditions, is what most often motivates the use of HMMs for analyzing individual animal movement data (see, e.g.\ \citealp{Morales2004}, \citealp{Patterson2009}, \citealp{McKellar2015}, \citealp{DeRuiter2016}). 

Figure \ref{elkfig} displays the estimated state-dependent step length and turning angle distributions, as well as the estimated effect of the distance to water covariate on the t.p.m. The state-dependent mean step lengths were estimated as $0.36$ and $3.53$ in states 1 and 2, respectively, and the associated estimated turning angle distributions indicate a tendency to reverse direction in state 1 and directional persistence in state 2. It seems reasonable to follow \citet{Morales2004} and label the two states ``encamped'' and ``exploratory''. The fitted model indicates that the probability of switching from ``encamped'' to ``exploratory'' was highest when close to water, while the probability of a reverse switch was highest when away from water (at the times the elks were located)---according to the fitted model, the ``exploratory'' state is hardly ever visited when at distances to water greater than three kilometres.

\begin{figure}[!htb]
\begin{center}
	\includegraphics[width=0.9
    \linewidth]{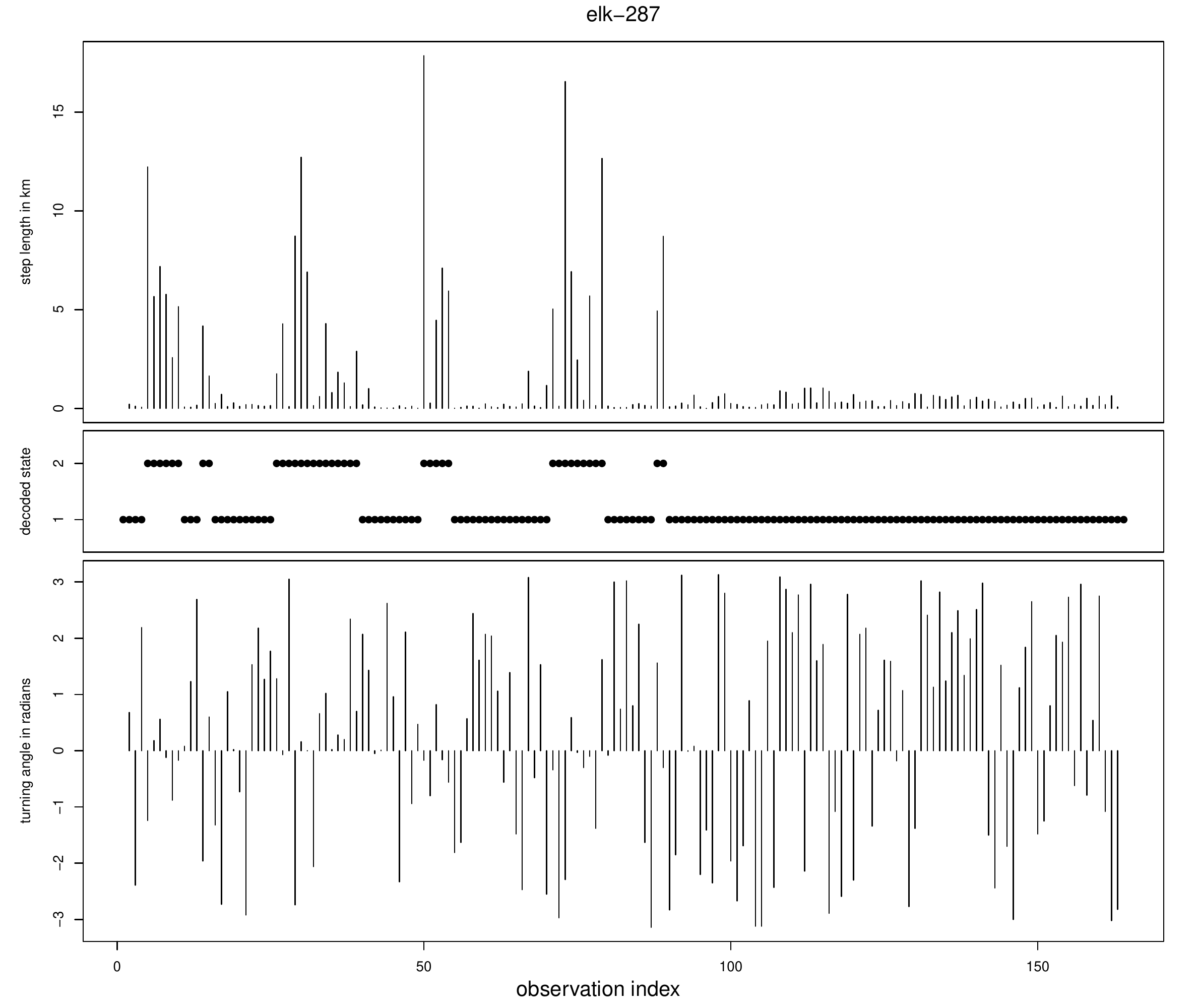}
	\caption{HMM applied to elk movement data: example Viterbi output (middle plot), together with corresponding step lengths (top plot) and turning angles (bottom plot), for one of the four elk.}
	\label{elkfig3}
\end{center}
\end{figure}

Figure \ref{elkfig3} displays, for elk-287, the state sequence that is most likely to have generated this elk's observations, under the fitted model. This sequence was obtained using the Viterbi algorithm. Notably, this elk was within 1 km of water for the first 89 days of observation (not shown in the figure)---during which the animal frequently switched between the ``encamped'' and the ``exploratory'' state---and $>1$ km away from water on days 90-164---during which the animal only occupied the ``encamped'' state.

\subsection{Limitations of the HMM framework}

The HMM framework is well suited to deal with animal positions that a) are observed at regular temporal spacings (and where the sampling unit needs to be meaningful with respect to the biological question of interest) and b) are observed with only negligible observation, or in this specific case, positional, error. 

Regarding a), it is straightforward to fit HMMs when data are missing at random on an otherwise regular grid. However, if the sampling protocol varies, or if observations are made essentially at random times, then the HMM machinery is not suitable. Discrete-time Markov chains are meaningless without reference to a sampling unit, and with irregular sampling there is also no obvious way to formulate for example a step length distribution that takes into account the amount of time passed between consecutive observations in a sensible way. (Note, however that there can be meaningful sampling units that do not involve a regular temporal grid, e.g.\ positions observed each time a marine mammal comes to the sea surface, such that the sampling is done on a dive-by-dive basis.) Continuous-time HMMs, with the underlying Markov process operating in continuous time, do exist (see, e.g. \citealp{Jackson2002}), but they are only suitable if the observed process has the ``snapshot'' property, such that the $k$-th observation, made at time $t_k$, depends only on the state active at time $t_k$ and not on the entire state trajectory over the interval $(t_{k-1},t_k]$. While this snapshot property is often naturally met in medical studies, this is generally not the case for the kind of movement data typically analyzed. 

Regarding b), when there is non-negligible measurement error in the locations---i.e.\ error that is too large relative to the step lengths and/or the question of interest to be ignored---then the basic HMM machinery is also not suitable. (If the locations are observed with error, then there is error in the step lengths and turning angles, and the way this error is generated does not allow for the use of say a simple convolution of step length and error distributions to be accommodated within the state-dependent process.) In the next section, we will first discuss how a class of models that is closely related to HMMs, namely state-space models (SSMs), can be utilized in order to deal with such positional error. At the end of that section, we will also return to a), the problem of irregularly spaced observations.

\section{State-space models: discrete time, measurement error}
\label{SSMs}

\subsection{Model formulation}

SSMs are doubly stochastic processes that are very closely related to HMMs. They have precisely the same dependence structure, with an observed time series such that any observation depends only on the current value of an underlying unobserved Markov state (or system) process. This can be generally expressed as 
\begin{eqnarray}\label{SSMform}
  \mathbf{z}_t =  f(\mathbf{z}_{t-1}, \varepsilon_t), \\
  \mathbf{y}_t =  g(\mathbf{z}_{t}, \eta_t), 
\end{eqnarray}
where the underlying latent state at time $t$ is given by $\mathbf{z}_t$ and the (typically noisy) observation of the latent state is $\mathbf{y}_t$. The functions $f(\cdot)$ and $g(\cdot)$ are the process and observations models, respectively, and $\varepsilon$ and $\eta$ are the process and observation errors, respectively. 

Some authors regard HMMs and SSMs as the same \citep{cap05}. However, the label HMM is usually used to indicate a model with a finite number of possible states, whereas in SSMs, the underlying state process typically takes continuous values and hence involves an infinite number of states. In the literature on movement modelling via state-switching processes, SSM approaches typically include both the (true) continuous movement metrics and the discrete states in the hidden component of the model, using the link to the observations to describe potential measurement error (see \citealp{Jonsen2005}, or \citealp{Patterson2008}). In contrast, in HMM approaches, as applied to GPS data, the measurement error is often assumed to be negligible, so that the hidden component of the model involves only the behavioural states, with the observed process giving the observed movement metrics, typically step lengths and turning angles. While this may be acceptable for GPS data, which are generally very precise, it will not be for other types of tag data such as that involving the use of  satellite tags or light-based geolocators.

\begin{figure}[!htb]
\begin{center}
	\includegraphics[trim={1.5cm, 20.5cm, 1cm, 0.5cm},clip,width=0.85\linewidth]{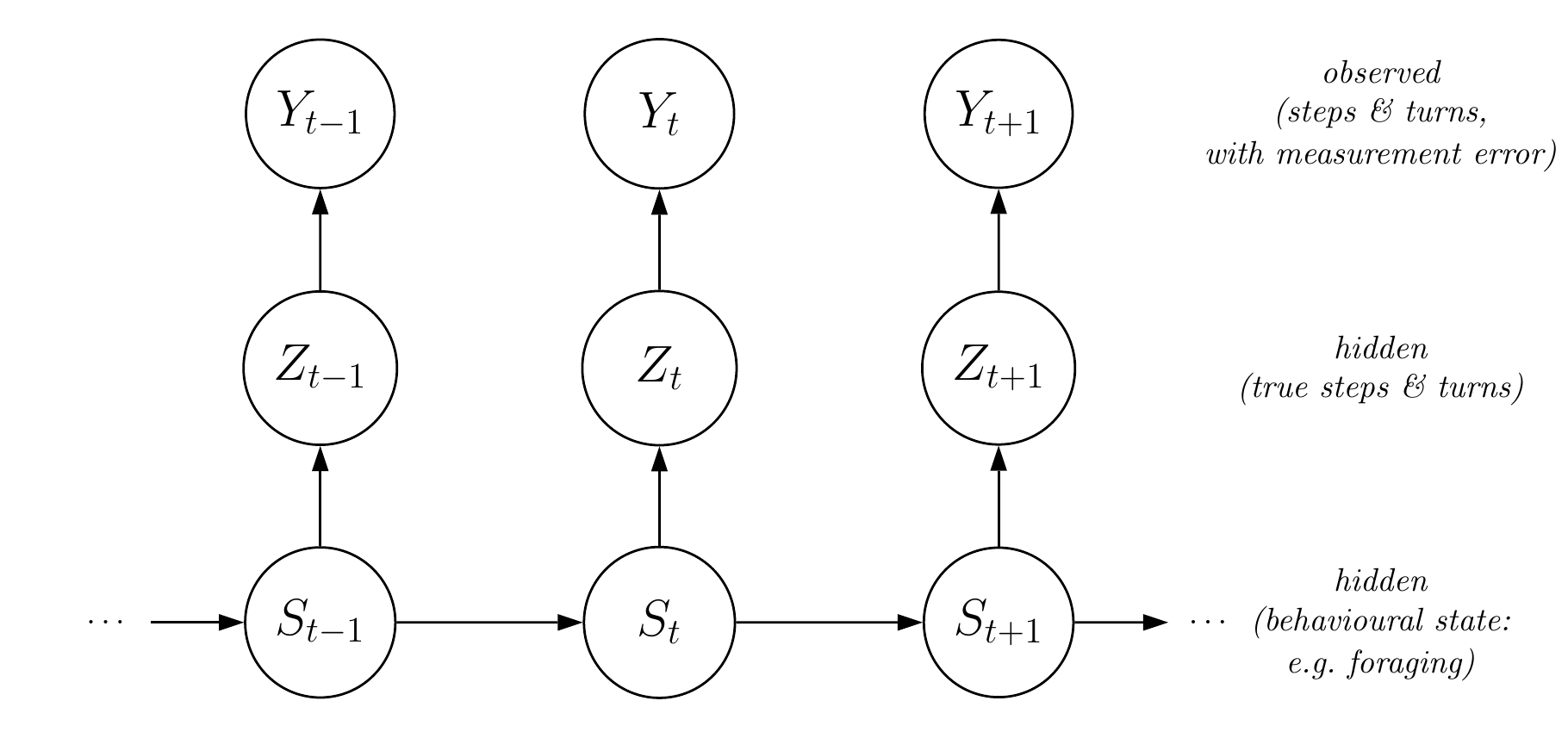}
	\caption{Structure of an SSM where observations are step lengths and turning angles, measured with error. This generalizes the model of Figure \ref{graph}. Note that while this structure is possible, it has not been implemented on real data and is likely to be difficult to apply (see text and Figure \ref{SSM2} for a more tractable structure).}
	\label{SSM1}
\end{center}
\end{figure}

\begin{figure}[!htb]
\begin{center}
    \includegraphics[trim={1.5cm, 20.5cm, 1cm, 0.5cm},clip,width=0.85\linewidth]{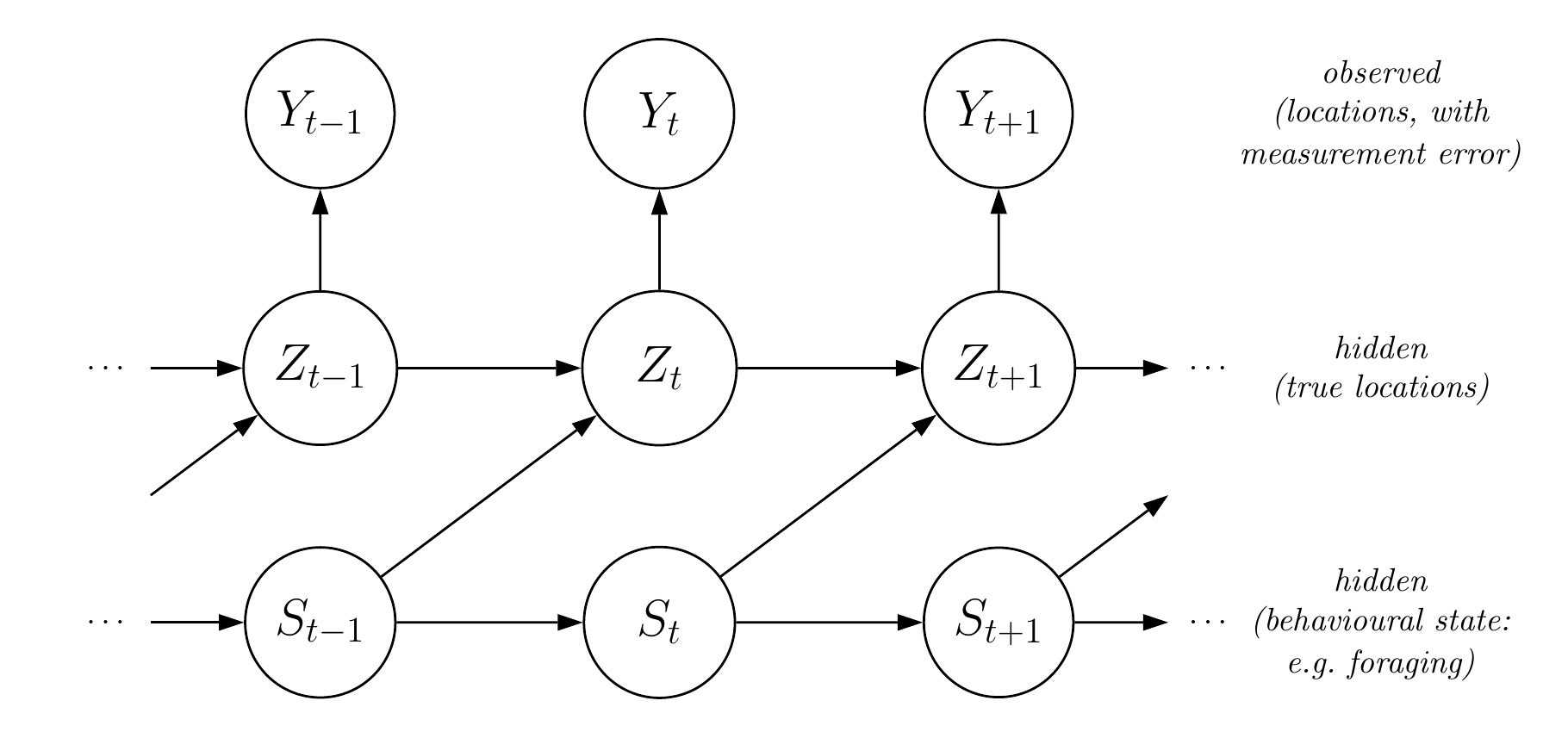}
	\caption{Structure of an SSM where observations are animal locations, measured with error.}
	\label{SSM2}
\end{center}
\end{figure}

We illustrate simple SSMs using Figures \ref{SSM1} and \ref{SSM2}.  Figure \ref{SSM1} shows an SSM that is a straightforward extension of the HMM in Figure \ref{graph} to allow for measurement error in the step lengths and turning angles. Here $s_t$ is the (discrete) state at time $t$, as before, and $\mathbf{z}_t$ is the vector of (continuous) state-dependent variables at time $t$, in this case the step length and turn angle.  However, unlike in Figure \ref{graph}, $\mathbf{z}_t$ is not observed directly, and so is modelled as a latent variable; together $\mathbf{x}_t=\{s_t, \mathbf{z}_t\}$ forms the hidden state at time $t$. Here, $\mathbf{y}_t$ is the observed step length and turn angle at time $t$, related to the true values through the observation process.  In practice, however, this model may not readily be applied: in particular errors in turn angle and step length may be correlated.  Nevertheless, for tags that measure speed (or acceleration) and bearing, rather than absolute location, models of this type, that include a component for measurement error in these quantities, may be useful (see \citealp{laplanche2015tracking}).  Typically, it is more natural for SSMs to model the true location of the animal as one of the hidden states, rather than the step length and angle; this provides a more direct link to the noisy observations on animal location that are provided by some types of tag (e.g. satellite/GPS tags) and it also allows for more explicit inferences about animal location.  This formulation is shown in Figure \ref{SSM2}: here $\mathbf{z}_t$ is true location and $\mathbf{y}_t$ is observed location.  Models that include elements of both formulations are, of course, possible---for example given a marine mammal tag that measures speed, orientation, depth and occasionally horizontal position (when the animal surfaces), one may envisage a 3D model like that of Figure \ref{SSM2} where $\mathbf{y}_t$ relates horizontal position and depth to true position $\mathbf{z}_t$, but with an additional observation model to link measured speed and orientation to change in true position.

In general, SSMs are much less mathematically tractable than HMMs.  The likelihood for HMMs, given in Eqn.\ (\ref{bruteforceL}), can be readily evaluated and maximized in the form of Eqn.\ (\ref{lik}).  By contrast, the likelihood for the model in Figure \ref{SSM1} is of the form
\begin{align}\label{SSM1L}
\nonumber \mathcal{L}^{\text{SSM1}}(\boldsymbol{\theta}) & = 
f(\mathbf{y}_1, \ldots, \mathbf{y}_T) \\
\nonumber & = 
\sum_{s_1=1}^N \int_{\mathbf{z}_1} \ldots \sum_{s_T=1}^N \int_{\mathbf{z}_T} f(\mathbf{y}_1, \ldots, \mathbf{y}_T | \mathbf{z}_1, \ldots, \mathbf{z}_T) f(\mathbf{z}_1, \ldots, \mathbf{z}_T | s_1, \ldots, s_T) f(s_1, \ldots, s_T) d\mathbf{z}_T \ldots d\mathbf{z}_1\\
& = \sum_{s_1=1}^N \int_{\mathbf{z}_1} \ldots \sum_{s_T=1}^N \int_{\mathbf{z}_T} \delta_{s_1} \prod_{t=1}^T f(\mathbf{y}_t | \mathbf{z}_t) f(\mathbf{z}_t | s_t)  \prod_{t=2}^T \gamma_{s_{t-1},s_t} d\mathbf{z}_T \ldots d\mathbf{z}_1 ,
\end{align}
and for the model in Figure \ref{SSM2} is similarly
\begin{align}\label{SSM2L}
\mathcal{L}^{\text{SSM2}}(\boldsymbol{\theta}) 
& = \sum_{s_1=1}^N \int_{\mathbf{z}_1} \ldots \sum_{s_T=1}^N \int_{\mathbf{z}_T} f(\mathbf{z}_1) \delta_{s_1} \prod_{t=1}^T f(\mathbf{y}_{t} | \mathbf{z}_{t}) \prod_{t=2}^T f(\mathbf{z}_{t} | \mathbf{z}_{t-1}, s_{t-1}) \gamma_{s_{t-1},s_t} d\mathbf{z}_T \ldots d\mathbf{z}_1 .
\end{align}
It is clear that evaluation of Eqns.\ (\ref{SSM1L}) or (\ref{SSM2L}) requires difficult (multi-dimensional) integrations; closed-form expressions are only available in special cases, where the integrands are of a particular form, such as the linear normal models of the Kalman filter, described below.  In other cases, model fitting is achieved using computer simulation-based methods within the Bayesian inference paradigm.  Two important exceptions are (1) the use of Laplace approximations to approximate the required integrals within a mixed-effects framework, and (2) discretization of the continuous states so that HMM machinery can be used.  In the next sections we describe briefly each of these approaches, starting with the Kalman filter, then discussing the mixed effects and discretization approximations, before considering two Bayesian inference methods: particle filters and Markov chain Monte Carlo.


\subsection{Kalman filters}
\label{Kalman}

The Kalman filter \citep{Kalman1960} is applicable in the special case of an SSM where the posterior distribution of the state, conditional on the previous observations, is analytically tractable. This tractability stems from two crucial assumptions: (1) that both the process and observation models are linear, and (2) that their respective error processes are Gaussian. The Kalman filter also is a recursive (and inherently Bayesian; see \citealp{wikle2007bayesian}) algorithm, updating state estimates while step-by-step traversing along the time series. Again analogous to the forward-backward algorithm in case of HMMs, the so-called Kalman smoother can be used to obtain state estimates given all observations and a fitted model. A good exposition of the Kalman filter is  given by \citet{Harvey1990}. Its development was a huge breakthrough in the application of state-space models to problems in engineering such as radar tracking, and has been applied to a vast number of problems in many fields. The Kalman filter and associated variants thereof have also been applied to animal movement data. In particular, one of the most widely used Kalman filters has been developed by Sibert, Nielsen and colleagues in a series of papers \citep{sibert2003horizontal,nielsen2006improving,nielsen2007state} which tackle the problem of estimating the position of animals (chiefly marine species) using ambient light data. \citet{Patterson2010} and \citet{johnson2008continuous} used Kalman filtering to satellite telemetry data from Service Argos. Indeed, Service Argos now employ Kalman filtering routinely to infer a most likely path from Doppler measurements from Platform Terminal Transmitter (PTT) devices \citep{lopez2014improving}. 


\subsection{Random effects approaches to SSMs}

The SSM formulation is a natural way to view the joint
problem of estimation of latent states given uncertain data---and
it fits very naturally to animal movement problems. The major barrier
to the widespread use of SSMs by ecologists are technical difficulties
in their implementation. Even the simplest linear SSMs are relatively complex for ``end users'', whose capacity to deploy SSMs may be limited by complexity of the necessary statistical machinery used to fit them. A relatively new approach to fitting SSMs offers analysts a more straightforward and flexible path to develop relatively flexible SSMs via mixed effects modelling.
In this section we closely follow the description given in \citet{fournier2012ad}. 

The usefulness of this approach comes through the ability to cast
the SSM as a more general hierarchical random-effects (or mixed-effects)
model. In this case the latent states are random effects, $\mathcal{Z}=\{\mathbf{z}_{1},\mathbf{z}_{2},\ldots,\mathbf{z}_{T}\}$.
A model for the data $Y_{\mathbf{T}}=\{\mathbf{y}_{1},\mathbf{y}_{2},\ldots,\mathbf{y}_{T}\}$,
conditional on the unobserved random effects, is given as $f_{Y_{\mathbf{T}}|\mathcal{Z}}(Y_{\mathbf{T}}|\mathcal{Z},\mathbf{\theta}_{y})$
along with a model of the unobserved random effects $f_{\mathcal{Z}}(\mathcal{Z}|\mathbf{\theta}_{z})$.
It is immediately obvious that these two components are equivalent
to the usual SSM components of the observation and process models.
From this, the joint density of both the latent states (random effects)
and observations conditional on the parameters $\mathbf{\theta}$ is
\begin{equation}
f_{\mathcal{Z},Y_{\mathbf{T}}}(\mathcal{Z},Y_{\mathbf{T}}|\mathbf{\theta})=f_{\mathcal{Z}}(\mathcal{Z}|\mathbf{\theta}_{z})\,f_{Y_{\mathbf{T}}|\mathcal{Z}}(y_{\mathbf{T}}|\mathcal{Z},\mathbf{\theta}_{y}).\label{eq:joint_density_RE}
\end{equation}
However, for estimating the parameters $\mathbf{\theta}=\{\mathbf{\mathbf{\theta}}_{z},\mathbf{\mathbf{\theta}}_{y}\}$
we require the marginal likelihood $\mathcal{L}_{M}(.)$. Obtaining this requires integrating
over the unobserved random effects

\begin{equation}
\mathcal{L}_{M}(\mathbf{\theta}|Y_{\mathbf{T}})=f_{Y_{\mathbf{T}}}(Y_{\mathbf{T}}|\theta)=\int_{\mathbf{R}^{\mathbf{T}}}f_{\mathcal{Z},Y_{\mathbf{T}}}(\mathcal{Z},Y_{\mathbf{T}}|\mathbf{\theta})d\mathcal{Z}.\label{eq:joint_marginal_llikelihood}
\end{equation}

In the software packages ADMB-RE \citep{fournier2012ad} and Template
model builder \citep{albertsen2015fast}, the  Laplace approximation is used to calculate a fast approximation of Equation (\ref{eq:joint_marginal_llikelihood}), carried out as follows: 
\begin{eqnarray*}
	\mathcal{L}_{M}(\theta;y) & = & \int L(\theta;z,y)dz\\
	& \approx & \int\exp\left(l(\theta;z,y)-\frac{1}{2}(z-\hat{z}_{\theta})^{\mathsf{T}}(-l''_{zz}(\theta;z,Y)|_{z=\hat{z}_{\theta}})(z-\hat{z}_{\theta})\right)dz\\
	& = & L(\theta;z,y)\int\exp\left(-\frac{1}{2}(z-\hat{z}_{\theta})^{\mathsf{T}}(-l''_{zz}(\theta;z,Y)|_{z=\hat{z}_{\theta}})(z-\hat{z}_{\theta})\right)dz\\
	& = & L(\theta;z,y)\cdot(2\pi)^{n/2}\cdot\text{det}\left(-l''_{zz}(\theta;z,Y)|_{z=\hat{z}_{\theta}})\right)^{-\frac{1}{2}},
\end{eqnarray*}
by taking the logarithm 
\begin{equation}
l_{M}(\theta;y)=l(\theta;z,y)-\frac{1}{2}\log\left(\text{det}\left(l''_{zz}(\theta;z,Y)|_{z=\hat{z}_{\theta}})\right)\right)+\frac{n}{2}\log(2\pi).\label{eq:neg_LLHD_LA}
\end{equation}

In ADMB-RE and TMB, automatic differentiation is used to compute the
Hessian matrix, $l''_{uu}(\theta;u,Y)|_{u=\hat{u}_{\theta}}$ of the
likelihood function, and minimization is done using standard numerical
methods. This avoids numerical approximation of Hessians and gradients,
which can lead to poor optimization performance through propagation
of errors in numerical differencing schemes. 

While these methods are well recognised in some disciplines, in particular in applied contexts such as fisheries science for fitting population dynamics models \citep{maunder2009comparison}, they are only starting to be more
widely used in general ecology and only very recently in movement ecology. A recent paper of \citet{albertsen2015fast} has applied these estimation methods to demonstrate estimation of a CRW
model which incorporates an Ornstein-Uhlenbeck process on the velocity
component. The authors provide an R package \texttt{argosTrack} which applies
these methods to Service Argos satellite telemetry data. This model is
essentially an extension of the \texttt{CRAWL} package \citep{Johnson2008},
which used Kalman filtering/smoothing of speed filtered Service Argos data.
However, by specifying the movement model within a mixed effects model
framework, the restriction of Gaussian error terms no longer applies,
and \citet{albertsen2015fast} demonstrate a model with $t$-distributed errors.
We feel the mixed effects approach to fitting SSMs in movement ecology
is an exciting and productive way forward as it offers a fast and
flexible method for estimating a range of movement models. Initially,
and like the \citet{albertsen2015fast} paper, the primary application will be
in constructing more flexible error correction filters. However, it
is likely that models with more ecologically interesting process dynamics
could be constructed within this modelling approach. This sort of
hierarchical state-space modelling has previously only been available
via complicated and bespoke modifications to Kalman filters (see e.g. \citealp{meinhold1989robustification}) or with MCMC sofware (e.g.\ WinBUGS, OpenBUGS, JAGS). These are discussed in papers by \citet{Jonsen2013} and \citet{Jonsen2006} (and see \citealp{pedersen2011estimation}, for a general comparison of techniques and software). 

As an illustrative example, consider a simple 2-dimensional problem. Let $\mathbf{z}_{t}=(z_{1,t},\,z_{2,t})$ be a 2-dimensional state variable (e.g.\ longitude and latitude, easting
and northing, etc.). 
\begin{eqnarray*}
	\mathbf{z}_{t} & = & \mathbf{z}_{t-1}+\eta_{t},\quad\eta_{t}\sim N(0,\Sigma_{z}),\\
	\mathbf{y}_{t} & = & \mathbf{z}_{t}+\varepsilon_{t},\quad\varepsilon_{t}\sim t_{\nu}(0,\Sigma_{y}).
\end{eqnarray*}

\begin{figure}[!htb]
\begin{center}
\includegraphics[width=0.7\textwidth]{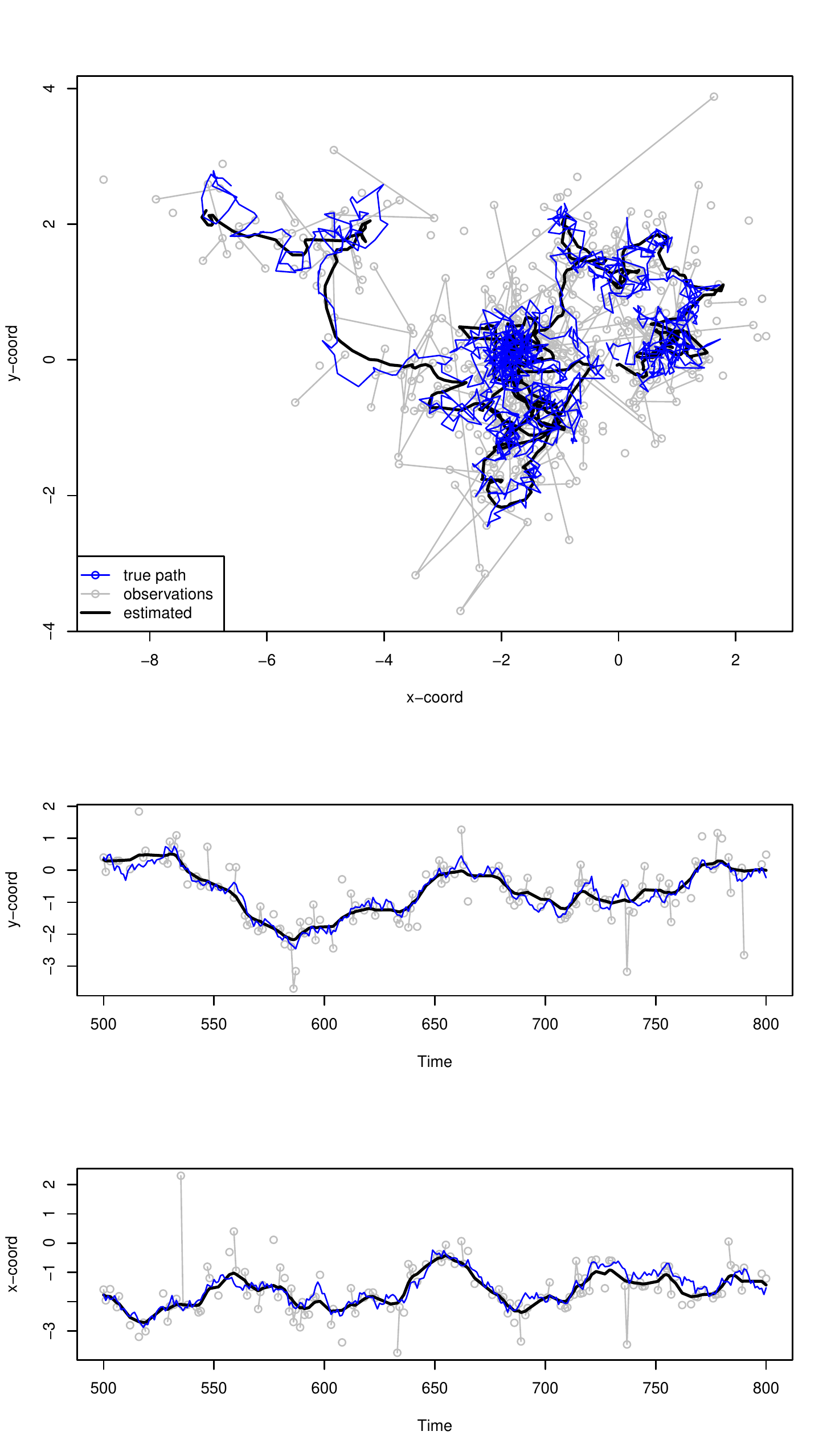}
	\caption{Estimated track from a random walk state space model fitted to simulated data with $t$-distributed errors and 40\% missing observations (N=2000 positions). Top panel: X-Y coordinates. Bottom panels: A section of 300 time steps from the simulated data shown in the top panel separated into X and Y components. In  all panels, lines between grey open circles denote consecutive observations; so if a joining line is absent this denotes an instance where an observation was missed. }
\label{SSMmixed}
\end{center}
\end{figure}

We simulate 2000 observations from a random walk with independent process errors $\Sigma_{z}=\mathbf{I}\sigma_{z}^{2}$. To demonstrate how the approach extends the canonical linear Gaussian case, we simulated a noisy heavy-tailed ($t$-distributed with degrees of freedom $\nu$) observation process $\Sigma_{y}=\mathbf{I}\sigma_{y}^{2}$ with 40\% of times having missing observations. Results are displayed in Figure \ref{SSMmixed}. This could not be accommodated using a standard Kalman filter. The package \texttt{TMB} was used to estimate the model parameters.

\subsection{Discretising space in SSMs}
\label{sec:discretisingspace}

In this approach, the continuous latent variables are finely discretized, so that the complexities of integrating over hidden states are reduced to a summation. The standard HMM machinery can then be applied as described in Section \ref{HMMs}. This is a powerful and underutilized approach which also has the advantages of being able to incorporate non-trivial spatial constraints such as animals having to avoid barriers to movement (i.e. water masses for ground dwelling animals that do not swim, or conversely land-avoidance in marine species). These methods were first demonstrated for geolocation from depth and temperature sensing tags by \citet{thygesen2009geolocating} and \citet{pedersen2008geolocation} and typically involve the use of data from sensors as well as (or instead of) noisy estimates of x-y position. The sensor data may be compared to spatial fields and a data likelihood can be generated for all states in the state space. From there the standard HMM routines can be applied. The case with Markov switching between diffusive and ballistic travel modes was shown by \citet{Pedersen2011}.  In all these studies, the transitions between latent states is governed by a PDE which is solved numerically to predict movement. A comparison of these methods to non-spatial / behaviour-only HMMs and switching CRWs fitted using MCMC, is given in \citet{Jonsen2013}. A limitation of spatial HMM approaches is that, given the large number of states which may need to be stored, computational aspects can be important. 

\subsection{Particle filters}

Particle filtering is a widely-used technique in computational statistics for making Bayesian inference from nonlinear SSMs where the emphasis is on ``online'' (i.e. real-time or near real-time) estimation of the underlying states---in our case, the true, but unknown, animal positions and behavioural states (where the application would be real-time tracking or location forecasting).  It is less commonly applied to offline or parameter estimation problems, although it can be used for both.  The nomenclature is not standardized in this area, and particle filtering is also referred to as sequential importance sampling and sequential Monte Carlo.  Like Markov chain Monte Carlo (MCMC, described in section \ref{BayesSSM}), particle filters can be used to make inference from very complex multi-state movement models.  Two strong advantages of particle filtering are (1) that it is very easy to set up the algorithm, requiring simply that one can simulate from the movement model and can evaluate the likelihood of observations given states, and (2) particle filtering can be fast to run compared with MCMC.  However, when parameter inference is required, as is commonly the case in movement modelling, these advantages typically disappear.  A practical disadvantage for practitioners is that general particle filtering software is not typically available, requiring custom-written code.

The starting point for both the particle filtering and MCMC approach to Bayesian inference on SSMs is to augment the set of model parameters to include additional auxiliary variables corresponding to the latent states--i.e. the true locations of individuals and their behavioural states---which we denote $\mathbf{x} = \{\mathbf{z}, \mathbf{s}\} = \{\mathbf{z}_1, \ldots, \mathbf{z}_T, s_1, \ldots, s_T\}$.  Further, we specify an initial latent state, $\mathbf{x}_0 = \{\mathbf{z}_0, s_0\}$, corresponding to the location and behavioural state at time 0.  Using Bayes theorem, the joint posterior distribution of the augmented model parameters, given the observed positions $\mathbf{y}$ can be written
\begin{eqnarray}
\pi(\mathbf{\theta},\mathbf{x}|\mathbf{y}) & \propto & g(\mathbf{y}|\mathbf{x},\mathbf{\theta}) f(\mathbf{x}|\mathbf{\theta}, \mathbf{x}_0) p(\mathbf{\theta}, \mathbf{x}_0) \nonumber \\
& = & \prod_{t=1}^T g(\mathbf{y}_t|\mathbf{x}_t,\mathbf{\theta}) f(\mathbf{x}_t|\mathbf{x}_{t-1},  \mathbf{\theta}) p(\mathbf{\theta}, \mathbf{x}_0), \label{postssm}
\end{eqnarray}
The (marginal) posterior distribution of the model parameters is obtained by integrating out the auxiliary variables. The necessary integration is, however, analytically intractable. 


Particle filtering (and MCMC) both work by generating samples from the posterior distribution $\pi(\mathbf{\theta},\mathbf{x}|\mathbf{y})$.  Inferences about model parameters, including latent states, is then made readily using techniques of Monte Carlo integration---for example, the posterior mean of $\mathbf{\theta}$, $\mathbf{z}$ or $s$ can be estimated from the mean of the sample values.  Both methods are described in many texts, but a good general introduction is  \citet{Liu2004}.  Introductory articles to particle filtering include \cite{Doucet2000}, \cite{Doucet2001}, and \cite{Arulampalam2002}, and a very basic application to a state-switching animal movement model is given by \cite{Patterson2008}.

Particle filtering generates independent samples from the posterior.  There are many variants, but the most basic, the bootstrap filter, can be viewed as an extension of importance sampling.  There, many replicate samples are simulated from a proposal distribution $q()$, and each is assigned an importance weight $w=\pi(\mathbf{\theta},\mathbf{x}|\mathbf{y})/q()$; the weighted samples can then be used to make inferences about the posterior distribution of interest.  These samples are called ``particles'' in particle filtering.  The bootstrap filter makes three extensions:
\begin{enumerate}
\item Advantage is taken of the Markovian nature of the model, so that the algorithm proceeds one time step at a time, starting by proposing values for time 1 based on an initial sample for time 0, calculating time-specific weights, $w_1$, and resampling (see next extension, below), then proposing values for time 2 based on the samples at time 1, calculating weights $w_2$ and resampling, etc. 
\item There is a resampling step at each time period, where the weighted particles are resampled with replacement with probability proportional to the weight, to yield an unweighted set of particles that can be used for inference.
\item The initial sample at time 0 comes from the prior $p(\theta, \mathbf{x}_0)$, and the proposal each time step is based on the process model $q_t =  f(\mathbf{x}_t|\mathbf{x}_{t-1},  \mathbf{\theta})$.  This results in a weight of a very simple form: $w_t = g(\mathbf{y}_t|\mathbf{x}_t,\mathbf{\theta})$, i.e. the observation process density.
\end{enumerate}

The filtering algorithm outlined above is also accompanied by a reverse algorithm, the particle smoother, that is exactly analogous to the backward step of the forward-backward algorithm in HMMs and the Kalman smoother in Kalman filtering.  This second stage is not relevant for online problems, but is typically applied in offline problems like post-processing animal movement data.

Hence, all that is required to implement the most basic particle filter (and smoother) is the ability to sample from the prior distributions of model parameters and latent states, to simulate realizations from the process model, and to evaluate the observation process density given values of the latent variables.

Unfortunately, the basic method can suffer from high Monte Carlo error, because resampling with replacement at each time step from the particles simulated at time 0 inevitably means that there are fewer and fewer of the unique ``ancestral'' particle remaining at each time period---a phenomenon known as ``particle depletion''.  This is not typically a serious problem for latent variables that are dynamic (i.e. time-varying) components of the process model, such as animal locations $\mathbf{z}_t$ and behavioural states $s_t$, because simulating stochastically from the proposal distribution (i.e. the process model) at each time step generates new diversity among the simulated particles.  However, for the static model parameters, $\mathbf{\theta}$, each time resampling is performed, fewer and fewer unique values from the original simulated set remain, resulting in an increasingly poor approximation to the posterior distribution.  One solution to this is to make the model parameters time varying, for example by having expected speed and turn angle evolve slowly over time according to a first order Markovian process.  This is the solution adopted by some authors, e.g.\ \cite{Dowd2011}.  Another solution is to extend the particle filtering algorithm to maintain diversity among particles in static parameter values, for example by resampling from kernel smoothed estimates of the joint posterior distribution of parameters or by introducing an MCMC step. An example of an MCMC step, applied after particle filtering in order to facilitate static parameter estimation, is \cite{Andersen2007}. Many other techniques are available (see review by \citealp{Kantas2015}); however, these methods tend to loose the advantages of simplicity and speed.  Perhaps because of this, or perhaps because of the absence of general software for particle filtering animal movement data, MCMC (as described in the next section) has been historically the more popular approach.



\subsection{Markov chain Monte Carlo}\label{BayesSSM}

Markov chain Monte Carlo (MCMC) is a very popular approach for obtaining inference on the model parameters within a Bayesian analysis by simulating (dependent) samples from the posterior distribution (Eqn.\ (\ref{postssm})). Standard easy-to-use computer packages exist which implement an MCMC algorithm for a given model, prior specification and associated data. The MCMC algorithm is performed within a closed ``black-box'', so that in-depth computational details of the algorithm are not required. The most widely used packages for movement models are BUGS and JAGS \citep{Lunn2000,Plummer2003}. \citet{Jonsen2005} provide BUGS code for fitting SSMs with multiple CRWs to animal movement data, which has been employed widely. However, we discuss below why such general software packages can perform poorly. Alternatively, bespoke MCMC computer codes can be written with complete control over the updating algorithm, permitting more general updating algorithms \citep{McClintock2012}. Typically this is a non-trivial endeavour. 

The general structure of the MCMC algorithm for sampling from the joint posterior distribution specified in Eqn.  (\ref{postssm}) is as follows. At each iteration of the MCMC algorithm, the model parameters, $\mathbf{\theta}$, and auxiliary variables, $\mathbf{x}$, are updated. For animal movement models, a mixture of single and block updates are generally used. Single updates are typically used for the model parameters, $\mathbf{\theta}$, and discrete behavioural state, $s_t$, at time $t$; and block updates for the true location of an individual at a given time $t$, $\mathbf{z}_t$ (i.e. the cartesian co-ordinates are updated simultaneously). Within the constructed Markov chain, each 
iteration involves cycling through each individual model parameter, behavioural state and location parameter (at time $t$) to update their values. 

For the behavioural state, $s_t$, the posterior conditional distribution is of Multinomial form with 1 trial and associated probability for state $i=1,\dots,N$,
\[
p_i = \frac{f(\mathbf{z}_t, s_t = i | \mathbf{z}_{t-1}, s_{t-1}) f(\mathbf{z}_{t+1}, s_{t+1} | \mathbf{z}_{t}, s_{t}=i)}{\sum_{j=1}^N f(\mathbf{z}_t, s_t = j | \mathbf{z}_{t-1}, s_{t-1}) f(\mathbf{z}_{t+1}, s_{t+1} | \mathbf{z}_{t}, s_{t}=j)}
\]
(for $t \ne 0,T$). Thus a Gibbs sampler can be implemented, such that at each iteration of the Markov chain the behavioural state is updated by simulating from the given Multinomial posterior conditional distribution. For the remaining parameters, the full posterior conditional distribution is of non-standard form so that a Metropolis-Hastings algorithm is used. For example, suppose that a given iteration of the Markov chain, the current location at time $t$ is $\mathbf{z}_t$ (for $t \ne 0, T$). Simulate the proposed value, $\mathbf{v} \sim q(\mathbf{v} | \mathbf{z}_t)$, where $q$ is the proposal distribution. For example, a common choice for the proposal distribution is a random walk, such that
$\mathbf{v} = \mathbf{z}_t + \mathbf{\epsilon}$, where $\mathbb{E}(\mathbf{\epsilon}) = \mathbf{0}
$.
The proposed value is accepted with probability, $\min(1,A)$, where
\begin{eqnarray*}
A & = & \frac{f(\mathbf{z}_{t+1}|\mathbf{v}, s_t, \mathbf{\theta}) f(\mathbf{v}|\mathbf{z}_{t-1},\mathbf{\theta})
q(\mathbf{z}_t | \mathbf{v})}
{f(\mathbf{z}_{t+1}|\mathbf{z}_t,\mathbf{\theta}) f(\mathbf{z}_t|\mathbf{z}_{t-1},\mathbf{\theta})q(\mathbf{v} | \mathbf{z}_t)},
\end{eqnarray*}
using the Markovian structure of SSMs. If the proposed value is rejected the chain remains in the same location state. The analogous Metropolis-Hastings updates are used for the remaining model parameters. See \citet{McClintock2012} for further discussion of such updating algorithms, including where transitions between states may be dependent on additional factors/covariates.

This form of updating leads to very high auto-correlation in the Markov chain for the simulated true location states, $\mathbf{z}_t$. This is a direct result of the high correlation between the location of an individual at time $t$, with their corresponding locations at time $t-1$ and $t+1$ (animals do not teleport). This can be immediately seen in the above acceptance probability for the Metropolis-Hastings algorithm for updating the location of an individual at time $t$---the acceptance probability is a function of the underlying density function for the movement of the individual in the intervals $[t-1,t]$ and $[t,t+1]$. This leads to generally very poor mixing within the Markov chain, and low effective sample sizes, so that large numbers of iterations are needed to obtain converged posterior estimates with small Monte Carlo error. Consequently, extensive posterior checking should be conducted to assess the convergence of the Markov chain, for example, using multiple Markov chains with over-dispersed initial values for the model parameters and auxiliary variables. For further discussion of these issues for general SSMs see, e.g. \citet{Fearnhead2011}.

Approaches have been proposed to improve the mixing of MCMC algorithms for SSMs. The most notable (and promising) approach considers a particle MCMC algorithm that combines particle filtering with MCMC \citep{Andrieu10}. The model parameters, $\mathbf{\theta}$, and discrete states, $\mathbf{s}$ are updated using an MCMC-type algorithm (for example a single-update Metropolis-Hastings within Gibbs algorithm) and the true locations, $\mathbf{z}$, updated using a particle filter. In addition, the behavioural states do not necessarily need to be imputed within the MCMC algorithm. Assuming first-order Markovian transitions between states, we can use the efficient HMM machinery to write down an explicit expression for the joint density (i.e. likelihood) of the true locations, given the model parameters (see Eqn. (\ref{lik})). In other words the behavioural states do not need to be treated as auxiliary variables and imputed within the MCMC algorithm.

Finally, we note that SSMs generally assume observations are recorded at a set of equally-spaced discrete-time intervals. In practice, irregularly spaced time steps can be forced into the regular time interval SSM framework by linearly interpolating the recorded location observations at the required time steps \citep{Jonsen2005,McClintock2012}. Discrete-time models have the advantage of accessible model-fitting tools (albeit potentially very inefficient) and immediately interpretable model parameters. However, issues arise, for example, if there is a mismatch between times between observation and the scale at which transitions occur between states. See \citet{McClintock2014} for an in-depth discussion of the issues of discretising time. An alternative and more natural approach in many situations (though at the expense of mathematical simplicity) is to consider continuous-time models, discussed in the next section.

\section{Diffusion models: continuous time}\label{DM}

Continuous-time modelling of movement almost always makes use of diffusion processes---
Markov processes with continuous sample paths. We distinguish two broad approaches: one is to build models from the limited selection of tractable diffusion models (Sections \ref{Brownian} to \ref{CTswitching}) and the other is to define models directly in terms of the stochastic differential equations that they satisfy (Section \ref{SDEs}). Our emphasis here is on the former, with animals switching between different movement modes---the direct equivalent in continuous time of the HMMs of Section \ref{HMMs}.

\subsection{Brownian motion}
\label{Brownian}
Brownian motion, or the Wiener process, the continuous-time version of a random walk model, is the simplest diffusion process, and its density function can be written down explicitly. If a Brownian motion $W(\cdot)$ starts at location $\mathbf{0}$ at time $0$, then 
\[
W(t) \sim  \normal(0,t\Sigma),
\]
where $\Sigma$ is a variance-covariance matrix (or, for one-dimensional movement, just a variance), most commonly with $\Sigma = \sigma^2 I_d$ (the isotropic or circular case, with $I_d$ the identity matrix of order $d$).
Brownian motion is an extremely simplistic model, representing an animal having no interaction with environment, and no directed or persistent movement. The model is   
of limited use in its own right, but becomes a useful component in switching models (Section \ref{Blackwell97}).

Given a sequence of observations 
$\{\mathbf{x}(t_j)\}$ in $d$ dimensions, the likelihood follows from the multivariate normal density;
\newcommand{\deltax}{\Delta\mathbf{x}}
\[
L(\Sigma|\{\mathbf{x}(t_j)\}) = \prod_j
{(2\pi)^{d/2}|\Delta t_j\Sigma|^{-1/2}
\exp(-(\deltax_j'(\delta t_j\Sigma)^{-1}\deltax_j)/2)},
\]
where $\deltax_j = \mathbf{x}(t_{j+1})-\mathbf{x}(t_j), ~\Delta t_j = t_{j+1}-t_j$.

Brownian motion is often used as a purely local model. 
\citet{Horne2007} suggest that movement between known locations can be estimated through the use of Brownian bridges; this possibility is also implicit in \citet{Blackwell1997,Blackwell2003}. 
\label{MathMeth:BB}
The Brownian bridge $(B(t), t_1\leq t\leq t_2)$ is the stochastic process that arises from 
Brownian motion $(W(t), t\geq 0)$ in which the state of the process is not only known at the start of the process, but also at the end~\citep{Anderson1996}. The Brownian bridge therefore describes 
Brownian motion conditioned on its state at these endpoints $t_1$ and $t_2$. 
%
With endpoints $B(t_1)=a$ and $B(t_2)=b$, the process has 
\begin{equation} \label{Eq:BB}
B(t) \sim \normal\left( a+\frac{t-t_1}{t_2-t_1}(b-a),\frac{\sigma^2(t_2-t)(t-t_1)}{t_2-t_1}\right),
\end{equation} 
for $t_1\leq t \leq t_2$~\citep{Anderson1996}.
Specifically, this kind of interpolation can be informative about 
an animal's utilization distribution---the marginal distribution of its location---and hence about its habitat use. 
%
Given two 
consecutive observations of location, it is assumed that the animal is moving according to Brownian motion and so the only movement parameter is the random volatility. 
The theory of Brownian bridges allows the likelihood of space use at any time between the two known locations to be easily evaluated, given the volatility parameter of movement. Furthermore, given a series of locations over time, disjoint Brownian bridges can be constructed between pairs of observations and the likelihood can be evaluated due to conditional independence.

\subsection{The Ornstein-Uhlenbeck position model }
\label{OU}

\label{MathMeth:OUIntro}
The Ornstein-Uhlenbeck (OU) process $(U(t), \ t \geq 0)$ is a stochastic process introduced by \citet{Uhlenbeck1930} 
as an improvement to methods for modelling the movement of particles---based on Brownian motion~\citep[see e.g.][]{Guttorp1995}. 
This stationary, Gaussian process is mean-reverting~\citep{Dunn1977}, and so has a tendency to drift towards its long-term mean. The equilibrium distribution of a particle following an OU process in $d$ dimensions is 
\begin{equation} \label{Eq:OUEquilib}
U(t) \sim \normal(\mu, \Lambda),
\end{equation}
where $U(t)$ and $\mu$ are $d$-dimensional vectors and $\Lambda$ is a $d \times d$ covariance matrix.
The conditional distribution of the process at a future point in time, given its current value, can be described as
\begin{equation} \label{Eq:OUCondit}
U(t+s) | U(s) \sim\normal(e^{Bt}U(s) + (1-e^{Bt})\mu, \ \Lambda - e^{Bt} \Lambda e^{B't}),
\end{equation}
where $U(t)$, $\mu$ and $\Lambda$ are as above and $B$ is a $d \times d$ stable matrix---that is, $e^{Bt}\rightarrow0$ as $t\rightarrow\infty$---so as to ensure a positive-definite covariance~\citep{Dunn1977}. It can therefore be seen that $\mu$ describes the centre of the process, with rate of attraction towards the centre controlled by $B$ and with random variation governed by $\Lambda$. 

The OU process is perhaps the simplest continuous-time model that is of use in its own right.
It arises from ecologists'  interest in learning about the home-range of an animal, the spatial range in which it performs its daily survival activities
~\citep{Borger2008}---often mathematically defined as the smallest geographical area in which the animal spends a fixed proportion of time~\citep{Jennrich1969}. Approaches to estimating the home range 
include that in \citet{Jennrich1969}, proposing that an animal's 
utilisation distribution (see Section \ref{Brownian})
can be represented as
a bivariate Normal distribution. 
%
This led to the first method for modelling animal positions in continuous time---given by \citet{Dunn1977}, who model the $(X,Y)$~co-ordinate positions of an animal by a 2-D OU process. The long-term position of the animal is described by the equilibrium distribution of an OU process--- see (\ref{Eq:OUEquilib}). Movement therefore has a random element to it, but the animal is ultimately attracted to a centre, and so has a well-defined home-range. Autocorrelation of successive observations is accounted for by the conditional distribution of the OU process---see (\ref{Eq:OUCondit}).

In the application of the OU process to animal movement, the matrix $B$ is often taken to be isotropic---uniform in all orientations---with $B=bI_d$ for $b<0$ to ensure stability. 
More general classes of $B$ may be used, but note that the class must be 
symmetric under rotation and reflection, 
otherwise there would be some significance placed on the co-ordinate system chosen~\citep{Dunn1977,Blackwell1997}. For example, the general diagonal case is not appropriate, for this reason.

Inference for the OU parameters governing movement in \citet{Dunn1977} is carried out by maximum likelihood methods. A difficulty presented by this method is the choice of likelihood for the initial observation. 
\citet{Dunn1977} explore two approaches. One is to ignore the information provided by the initial observation, effectively conditioning on that observation; this is the approach used most widely in movement analysis. The other is to use the tractability of the equilibrium distribution for the animal's position and add a likelihood term assuming that the initial observation comes from that equilibrium distribution. They discuss the implications  in terms of statistical information, and the relationship with the actual sampling scheme for the data.

The OU process addresses the problem of autocorrelation of position, meaning high frequency `bursts' of observations can be modelled. The OU process however, will always result in an estimate of home-range being elliptical and unimodal. For some animals and habitats this will clearly not be an appropriate assumption~\citep{Blackwell1997}.
Again, this limits the usefulness of the model on its own, but it is an important component in constructing more realistic models (Section \ref{CTswitching}).

\subsection{The Ornstein-Uhlenbeck velocity model}
\label{Velocity}
The 
persistent movement 
modelled using 
correlated random walks (Section \ref{HMMformulation} can be extended to a continuous-time framework. One such approach is given by \citet{Johnson2008} and applied to data from northern fur seals in \citet{Kuhn2009}. 
\citet{Johnson2008} model positions over time indirectly, by formulating a model in terms of velocity---the instantaneous rate of change of location. The 
behaviour of the velocity vector over time is then described by a bivariate OU process---in practice \citet{Johnson2008}
use two independent 1-D OU processes.
The persistence assumption on how animals move is thus incorporated as a result of the autocorrelation of the OU process.

The location of the animal at any time, $t$, can then be found by integrating the velocity process up to time $t$. This results in the location process no longer being Markovian---as in the OU position model above---as it depends on the entire velocity process prior to time $t$. However, the combined process of position and velocity \emph{is} Markovian in this model. Observation error in position is incorporated into \citet{Johnson2008} via a SSM with Gaussian distributed errors and extended in \citet{albertsen2015fast} to allow for non-Gaussian errors.

Statistical inference for all unknown parameters in \citet{Johnson2008} is carried out using maximum likelihood techniques.  Kalman filtering---see Section \ref{Kalman}---is used to find these maximum likelihood estimates, along with prediction intervals for the velocity and location of the animal at unobserved times. In \citet{albertsen2015fast} inference is carried out via the Laplace approximation using the R package \texttt{TMB}.  

\subsection{Modelling switching behaviour in continuous time}
\label{CTswitching}
\label{Blackwell97}
\citet{Blackwell1997} suggests an extension to the Brownian and OU models in order to allow for behavioural `switching'. As in the HMMs of Section \ref{HMMs}, it is assumed that at any point in time an animal exhibits one of a finite set of behavioural states. The process describing the behavioural state of the animal is assumed to follow a continuous-time Markov process.
%
The animal's movement is modelled in the same way as in \citet{Dunn1977} by an OU process. The OU process parameters however, are dependent on the behaviour process; when the animal is in behavioural state $i$ it moves according to an OU process with the parameters $\mu_i$, $\Lambda_i$, $B_i$~\citep{Blackwell1997}---see (~\ref{Eq:OUCondit}). Brownian motion can be recovered as a limiting special case.

The Markov process $M(t)$ taking values from a finite state-space of size $N$ can be fully described by its generator matrix, $G=\lbrace g_{ij} \rbrace$ for $i,j=1,\ldots,N$. The values $g_{ij},~i\neq j$ describe the infinitesimal transition rate from state $i$ to state $j$;
the rate of transitions out of state $i$ is given by $-g_{ii}$. The process can therefore be thought of as being in a state $i$ for a length of time exponentially distributed with mean 
$-{1}/{g_{ii}}$, 
and then `switching' to another state $j$ with probability 
for $i \neq j$. 
An alternative parametrisation~\citep{Guttorp1995} of the process is therefore given by the transition rates out of each state, $\boldsymbol{\lambda} =\lbrace \lambda_i \rbrace = \lbrace -g_{ii} \rbrace$, and the set of jump probabilities $Q=\lbrace q_{ij}\rbrace = \lbrace \frac{g_{ij}}{-g_{ii}} \rbrace$ for $i \neq j$.

Given observed data of a continuous-time Markov chain, sufficient statistics for the process parameters $\boldsymbol
{\lambda}$ and $Q$ are given by $\mathbf{t}=\lbrace t_i \rbrace$, the total observed times spent in each state $i$, and $\mathbf{n}=\lbrace n_{ij} \rbrace$, the numbers of observed transitions from state $i$ to state $j$~\citep{Guttorp1995}. The likelihood for $\boldsymbol{\lambda}$ and $Q$ is then given by
\begin{equation} \label{Eq:MarkLik}
L(\boldsymbol{\lambda}, Q ; \boldsymbol{t}, \boldsymbol{n}) = e^{\sum_i \lambda_i t_i} \prod_{i \neq j} (q_{ij}\lambda_i)^{n_{ij}}.
\end{equation}

Statistical inference for 
these models is given in \citet{Blackwell2003}, 
applied to positional data with known behavioural states at each observation time. Inference is more complicated than in the \citet{Dunn1977} case as the conditional distribution of the animal's position---given an earlier position in time---depends on the complete behaviour process between these two times. This entire behaviour process however, is unknown. 
The approach taken by \citet{Blackwell2003} treats the behaviour process as ``missing'' data and uses Markov chain Monte Carlo (MCMC) techniques. Quantities of interest are split into three groups and a hybrid MCMC is carried out, where posterior distributions are sampled from each group separately, using Gibbs sampling techniques. The three groups are the `missing' complete behaviour process, the behaviour process parameters and the movement process parameters. 
\cite{Blackwell1997,Blackwell2003} assumes that the behaviour process is independent of the geographical position of the animal.
\citet{Harris2013} describe spatially heterogeneous extensions of these models, where movement and behaviour may depend on the discrete spatial region in which an animal is located at a given instant. 
\citet{BlackwellExact} 
give a method of Bayesian inference for models where switching probabilities may vary with spatial location, in either discrete or continuous form, and with time; behaviour is generally taken to be unknown and is reconstructed as part of the MCMC algorithm. 

It is important to note that, while it is convenient to refer to the ``behaviour process'', the behavioural state potentially has the same limitations as in the HMMs of Section \ref{HMMs}; that is, the state may reflect a statistical description of movement rather than necessarily being ``behaviour'' in true biological sense. 
See Section \ref{states} for further discussion.

It is also worth pointing out that a `behaviour' here simply refers to a set of parameter values, and so different behavioural states may simply represent, for example, similar kinds of movement centred on different points of attraction. Combined with dependence of the switching probabilities on location, this means that these models can represent quite varied interactions with spatially complex environments. See \citet{Harris2013} for a range of examples, and \citet{BlackwellExact} for statistical analysis (using the methods outlined in Section \ref{exact}) of movement in a habitat known from satellite imaging.

\subsubsection{Simulation of a switching diffusion process}
\label{simulateMP}
To understand the basis of the estimation approaches described previously, it is instructive to consider how to simulate from a switching diffusion process. 

Consider a Markov process $M(t)$ on the finite set of states $S=\{1,\ldots,N\}$ with underlying generator matrix $G=\lbrace g_{ij}\rbrace$. Starting from some initial state $s_0$ at time $t_0$, 
this process can be simulated forward until some `end time', $T$, using the characterisation given above. This simulation approach can be extended to incorporate movement based on switching between some set of diffusion models. 
Algorithm~\ref{Alg:SwitchDiff}
gives details
in the form of pseudocode. Simply ignoring the locations gives a simulation of the Markov behaviour process.

\begin{algorithm}
\setstretch{1.4}
\begin{algorithmic}[1]
		\STATE Set the current time, $t\gets t_0$, state, $s \gets s_0$ and location $x\gets x_0$
        \STATE Generate $t^* \sim \textrm{Exponential}(-g_{ss})$
		\WHILE {$t+t^*< T$} 
        \STATE Generate $x^*$ from movement model $s$, starting at $x$, run for duration $t^*$
        \STATE Generate $s^*$ from the discrete distribution over $S\setminus s$ with probabilities ${-g_{s\cdot}}/{g_{ss}}$
        \STATE Update $t\gets t+t^*, s\gets s^*, x\gets x^*$ and store
        \STATE Generate $t^* \sim \textrm{Exponential}(-g_{ss})$
		\ENDWHILE
        \STATE Generate $x^*$ from movement model $s$, starting at $x$, run for duration $T-t$
        \STATE Store final state $T, s, X^*$
	\end{algorithmic}
\caption{Simulating a Switching Diffusion model with behaviour following a Markov Process}
	\label{Alg:SwitchDiff}
\end{algorithm}

\begin{figure}[hbt]
\begin{center}
	\includegraphics[width=0.75\linewidth]{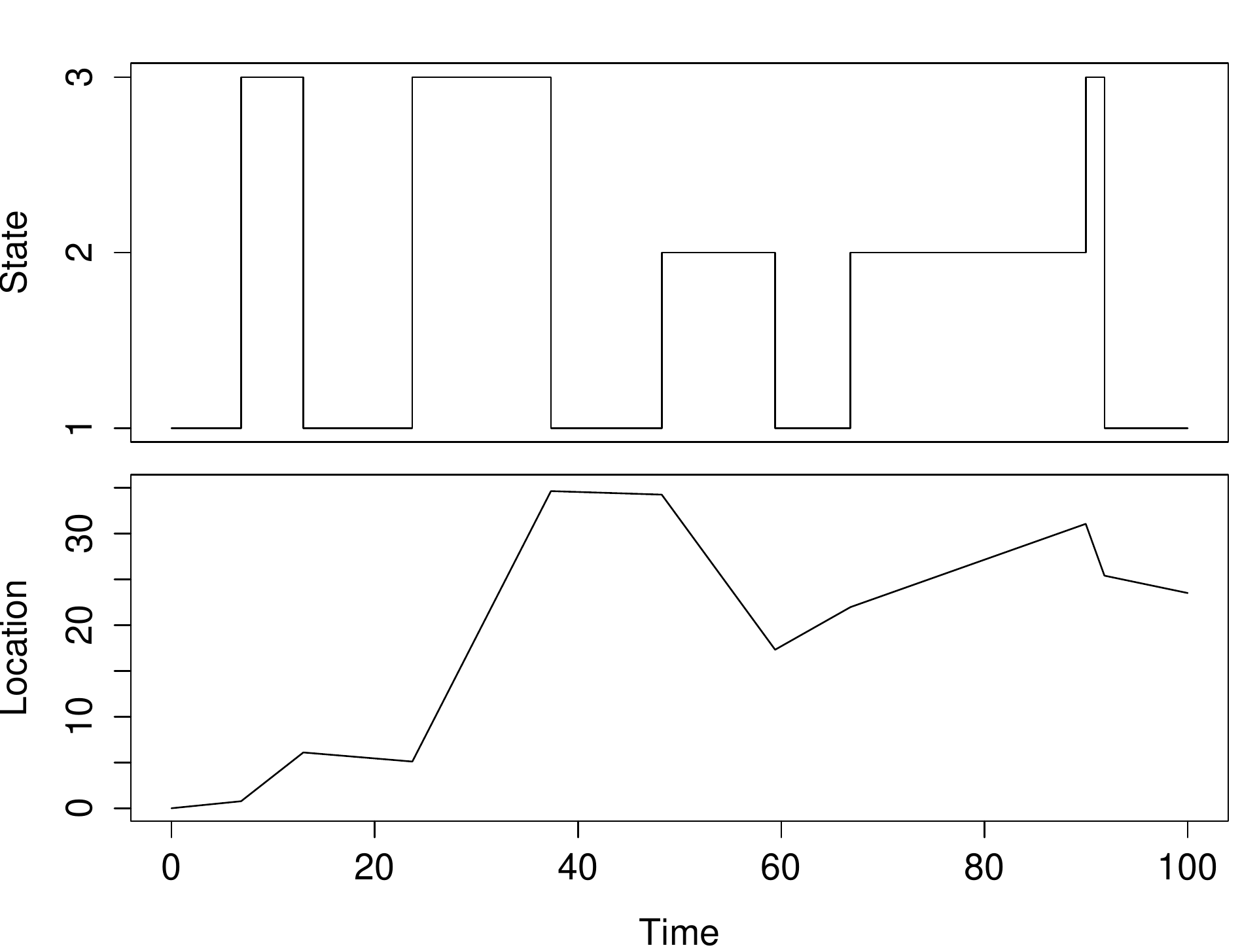}
	\caption{A simulated realisation of the Markov process starting with an initial state of 1 over 100~time-units and following the generator matrix given in Equation~\ref{Eq:GenMatrix}.}
	\label{Fig:SimMarkProcess}
\end{center}
\end{figure}

Assume a movement model in one dimension following Brownian motion with volatility $V=(0.01,0.1,1)$, determined by a Markov process on 3~states with generator matrix
\begin{equation} \label{Eq:GenMatrix}
G=\begin{pmatrix}
-0.10 & 0.04 & 0.06 \\
0.025 & -0.05 & 0.025 \\
0.20 & 0.00 & -0.20
\end{pmatrix}.
\end{equation}
A simulated realisation of this behavioural process and the corresponding locations at behavioural switching times is shown in Figure~\ref{Fig:SimMarkProcess} for 100~time-units after starting in state~1 and at the origin. Note that only locations at the times of switches are shown, as generated by the simulation. The Markov nature of the movement between switches means that additional points of the trajectory can readily be filled in using the idea of a Brownian bridge.

\subsubsection{Inference for switching diffusions}
\label{exact}
\begin{figure}[!htbp]
\begin{center}
\includegraphics[scale=0.55]{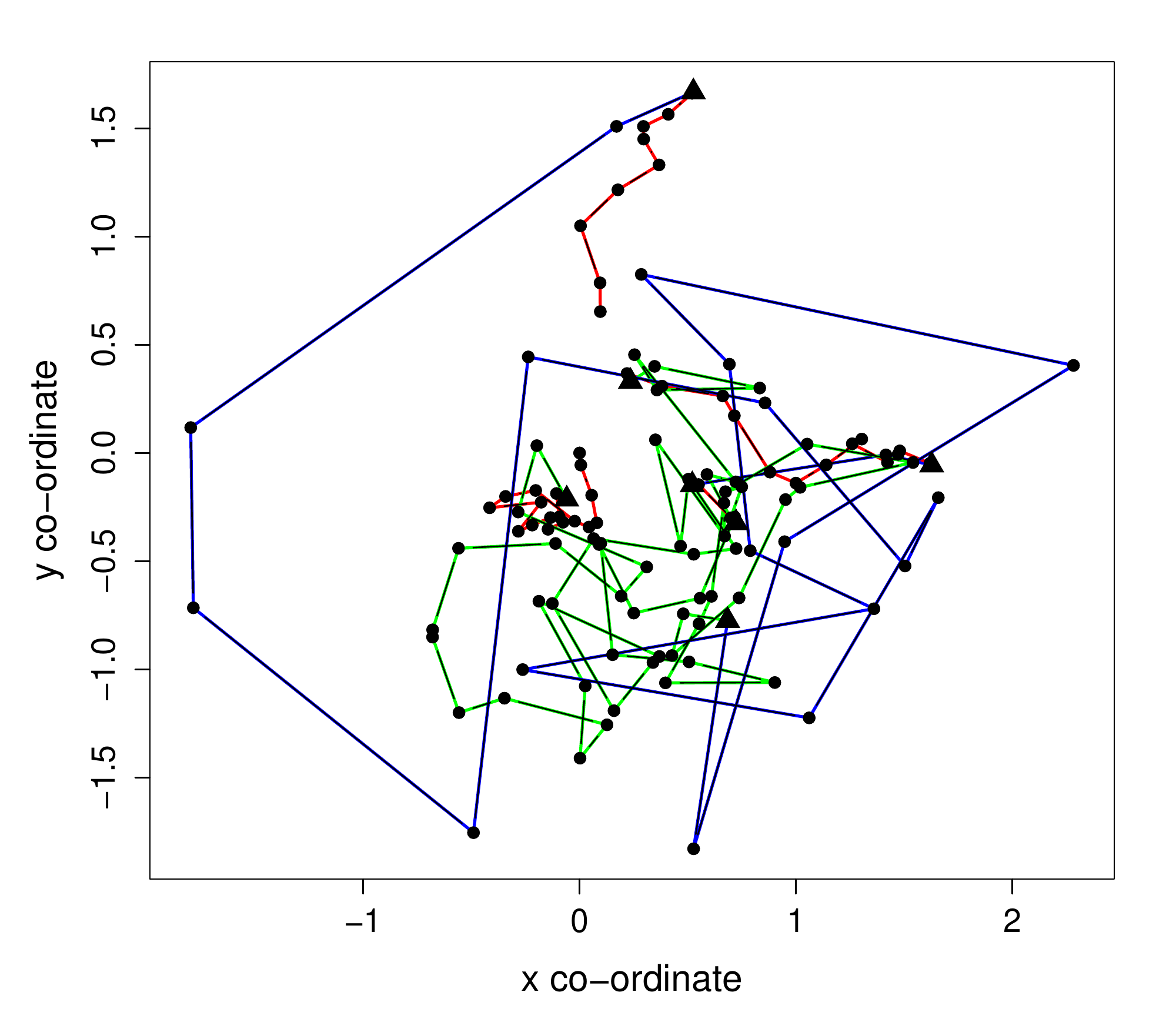}
\caption{Simulated trajectory from the switching OU model. Lines indicating movement in states 1, 2 and 3 are shown in red, green and blue respectively. Locations when behaviour switches are shown as triangles, other locations as circles.} 
\label{simData}
\end{center}
\end{figure}

\begin{figure}[htbp]
\begin{center}
\includegraphics[scale=0.55]{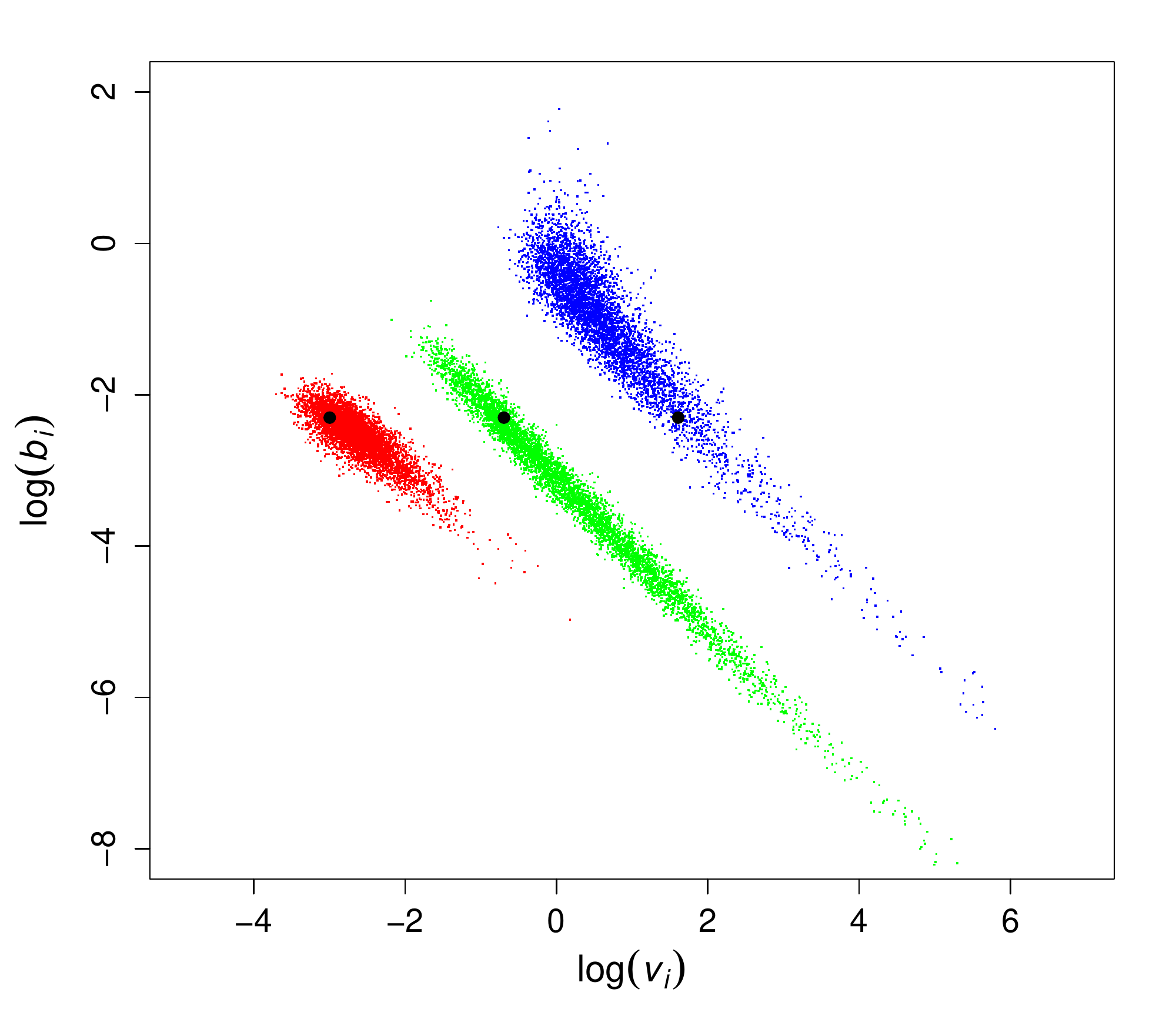}
\caption{Posterior distributions for parameters of the three behavioural states in the switching OU example. Parameters for states 1, 2 and 3 are show in red, green and blue respectively. True values are shown as black disks.} 
\label{logVB}
\end{center}
\end{figure}

\begin{figure}[htbp]
\begin{center}
\includegraphics[scale=0.5]{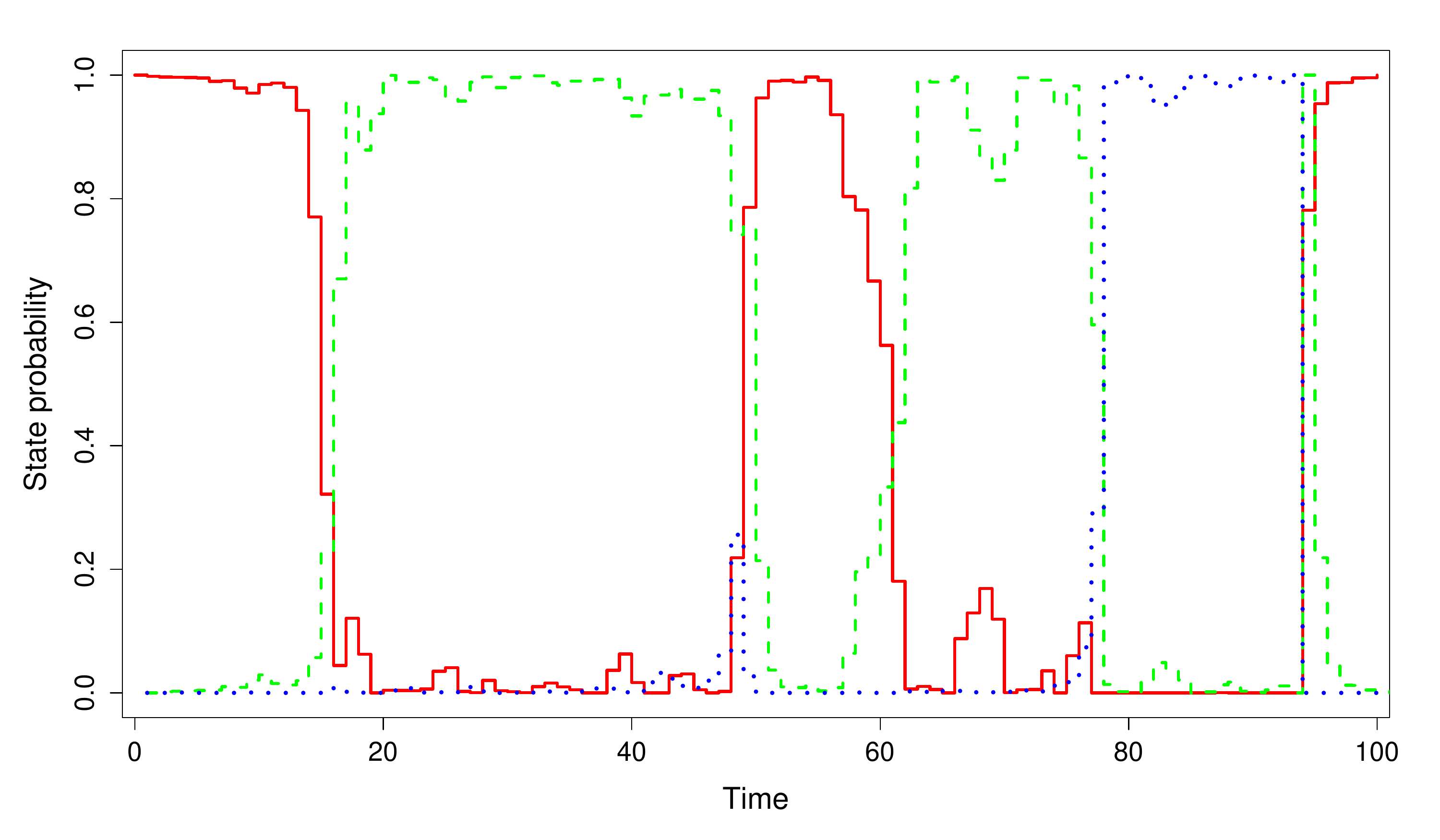}
\caption{Posterior probabilities for behavioural states, at the times of observations, in the switching OU example. Probabilities for states 1, 2 and 3 are show as solid red, dashed green and dotted blue curves respectively} 
\label{StateProb}
\end{center}
\end{figure}

\begin{figure}[htbp]
\begin{center}
\includegraphics[scale=0.5]{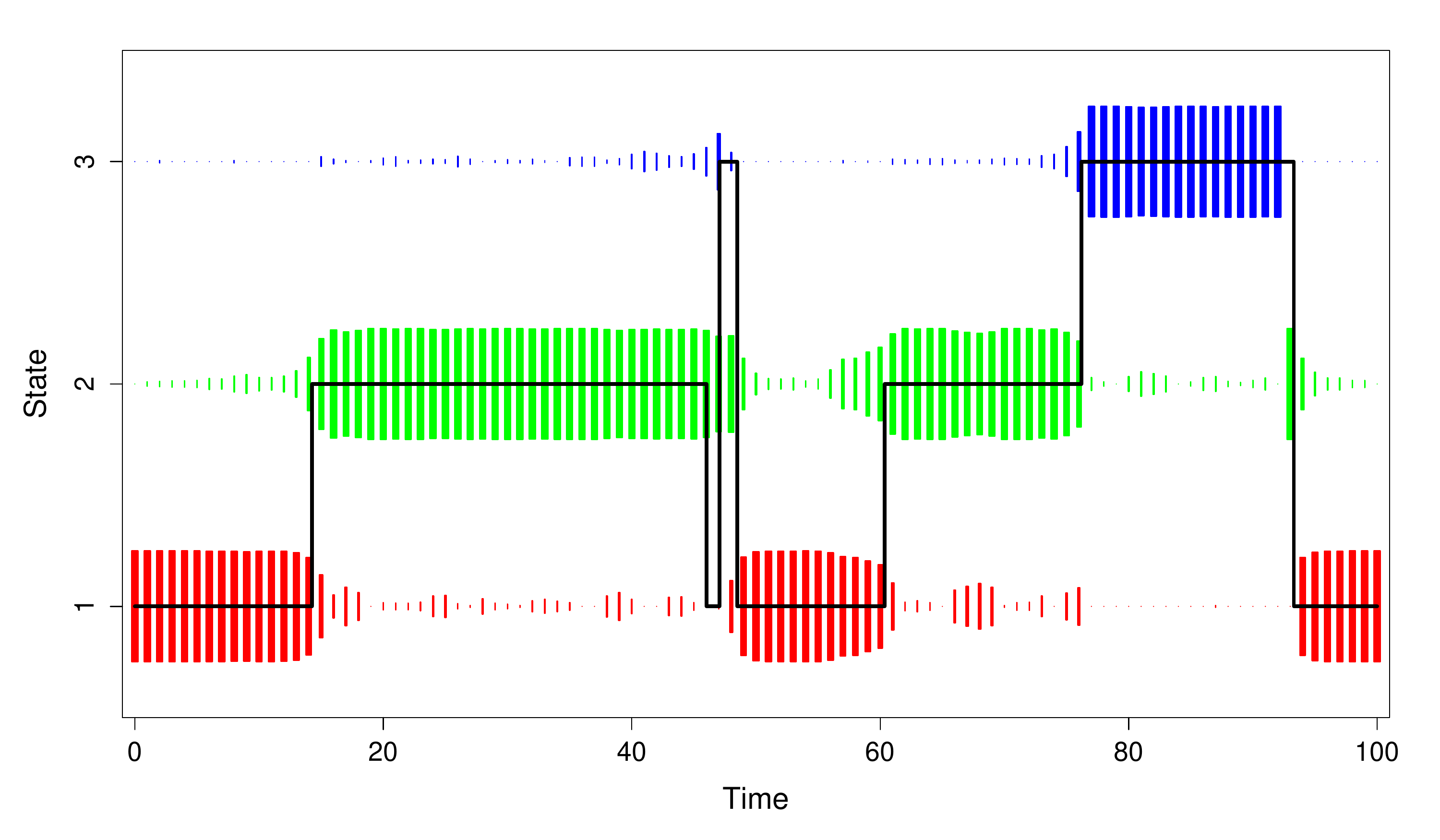}
\caption{Posterior probabilities for behavioural states, at the times of observations, in the switching OU example. Probabilities for states 1, 2 and 3 are indicated as areas of filled red, green and blue rectangles respectively; the largest rectangle corresponds to a probability of 1. The true state (known at all times, since the data are simulated) is shown by the solid curve} 
\label{StateBoxes}
\end{center}
\end{figure}

As an example of inference for continuous-time models, we again look at simulated data, from a model switching between three behavioural states according to a Markov process defined by the generator in \ref{Eq:GenMatrix}. Rather than the 1-d Brownian motion in Section \ref{simulateMP}, we consider a more realistic model with 2-d movement in each state following an OU process (for position, not velocity). This model is an example of those discussed by \cite{Blackwell1997, Blackwell2003}
and so can be fitted using the MCMC methods described there;
here we use a spatially homogeneous special case of the more general method, and code, in \citet{BlackwellExact}.
Figure \ref{simData} shows the simulated trajectory from the model.
Our prior distribution of the parameters takes the states to be ordered in terms of the derived quantity $\Phi$, the variance on the right hand of  \ref{Eq:OUCondit} with $t=1$, for each state. This enables us to avoid problems of label-switching, since behaviour is not observed and there is no inherent meaning to the labelling of states. 
Figure \ref{logVB} shows the posterior distributions, in the form of MCMC samples, for the two key parameters controlling the dynamics in this model, for each behavioural state.
Figures \ref{StateProb} and \ref{StateBoxes} show two ways of visualising the results of the reconstruction of states in this example,  based on an MCMC run of 1 million iterations. Fig.\ \ref{StateProb} shows the posterior probabilities (vertical axis)  of the true state taking each of the three  possible values, at the time of each observation. Fig.\ \ref{StateBoxes} shows the same probabilities as areas of the solid rectangles, with the true trajectory of the process through the three states superimposed as a solid line.

\subsubsection{Computation for switching diffusions}
The computation for the method of Section \ref{exact} is very time consuming, as it involves MCMC sampling of behavioural trajectories, conditional on data, which have varying numbers of changepoints and can typically only be updated a short segment at a time. As such, these methods are limited in their application at present, and are certainly not yet feasible for data-sets with  very large numbers of observations, or large numbers of individuals. 
However, high computational cost is not inherent in these models; improving the algorithms and their implementation, and developing fast and efficient approximations, is a very fast-moving area of research, making use of computational ideas from other strands of movement modelling, broader advances in Bayesian computation, and techniques from stochastic modelling generally. We include this approach since we believe it has a place in movement modelling which will only increase.

\subsection{Stochastic differential equations}
\label{SDEs}
The diffusion models described so far 
are tractable because they are linear and Gaussian. A more flexible modelling approach is to describe movement within a state implicitly, in terms of a stochastic differential equation (SDE)
. 

Brownian motion remains a key component in defining such models. A general SDE can be written as 
\begin{equation} \label{Eq:SDE}
dX(t) = A(t,X(t))dt + B(t,X(t)) dW(t),
\end{equation}
where $W(t)$ is Brownian motion. Discussion of the formal meaning of such equations is beyond the scope of this paper; we look briefly at an intuitive level. The simplest non-trivial example is the OU process, as above, which in one dimension can be derived as the solution to the stochastic differential equation
\begin{equation} \label{Eq:OU}
dU(t) = -a(U(t) -\mu)dt + \sigma dW(t),
\end{equation}
since the attraction towards the long-term centre $\mu$ is linear.

SDEs can describe much more flexible movement models, generally at the expense of computational, and hence statistical, tractability. For example, \cite{BrillingerStewart}, \cite{Brillinger}, \cite{Preisler2004} and \cite{Preisler2013} all consider the case where the SDE derives from a potential function, by taking $A(t,X(t))$ in \ref{Eq:SDE} to be minus the gradient of the potential function, representing an animal's attraction to or avoidance of a particular point, line or region in a completely general way. \cite{BrillingerStewart} and \cite{Brillinger} also use models for movement that are defined through SDEs incorporating spherical geometry, allowing a natural representation of long-range migration along ``great circle'' routes. 

In all these cases, some element of approximation is needed to fit these models. Typically, a normal approximation to the movement over each time-step is used 
\[
\mathbf{x(t_{j+1})} \sim \normal\left(
\mathbf{x(t_{j})} + A(\mathbf{t_{j}},\mathbf{x(t_{j})})\Delta t_j,
B(\mathbf{t_{j}},\mathbf{x(t_{j})})\Delta t_j\right).
\]
An approximate likelihood can then be derived and maximised; the quality of the approximation obviously depends on the frequency of the data compared with the rates at which $A(\cdot,\cdot), B(\cdot,\cdot)$ vary. More sophisticated approaches to inference from SDEs are available, but seem to be rarely used in a movement context, 
because of the extra computational cost, particularly in the presence of measurement error.

All the SDE models above take the state of the process to be the animal's location. Recent work~\citep{Parton2016} explores a different representation, with the animal's bearing and speed following SDEs to give a continuous-time version of the step-and-turn models described earlier. Implementation involves reconstructing the animal's path using MCMC on a finer time-scale than that of the observations, avoiding the arbitrariness of the latter at some computational cost.

\section{Discussion}
\label{discuss}

Given the numerous approaches that have been proposed for the analysis of animal movement data, our review is necessarily myopic in order to avoid superficiality. Thus, our review does not cover all relevant existing approaches to animal movement modelling, instead focusing on what we believe to be a few of the key tools for conducting meaningful biological inference from movement data collected at a relatively fine temporal scale. The rationale behind this was to provide researchers working on this type of data with a concise overview of the basic toolbox of which we think they ought to be aware of. We will organize the discussion in the same spirit, not attempting to cover a wide range of topics, including for example the various future directions of research on animal movement. Instead, we focus on the discussion of what we consider to be crucial, but sometimes neglected, issues concerning good practice in animal movement modelling. 

\subsection{Formulation of study aims and study design}

An area that receives very little attention is that of design of movement studies. This in itself may cover a variety of aspects. One example is estimation of data-throughput- the amount of data we expect to retrieve from a single instrument. Another is how to optimize data returns given constraints of bandwidth limitations (e.g.\ from satellite tags) against expected longevity. Various aspects of this were tackled by \citet{patterson2011designing} with regard to Service Argos. In that work, a model of failure rates was fitted to previous tag deployment data, and a simple model of transmission schedules dependent on the location on the globe were presented. \citet{musyl2011performance} also examined failure rates of similar instruments given various aspects of the subject animal and tagging protocol. \citet{breed2011electronic} examined how the choice of duty-cycling in satellite tags drastically affects the results from Bayesian switching models (in the spirit of \citealp{Jonsen2005}). Another study, by \citet{bidder2014risky}, examined the use of engineering approaches to analysis of failure events in biotelemetry. Indeed, many of these are ``engineering'' issues, rather than statistical or ecological issues, although they are crucially important to determining what ecological inferences may safely be drawn from a particular data set. 

More broadly, we are not currently aware of a study that seeks to ask ``How many tags on species X are required to estimate an effect Y?'' Here, Y could be the influence of habitat type on movement behaviour or characterising sex-specific movement rates, or a multitude of other questions. A possible exception to this is the work by \citet{pagendam2011optimal}, which looks at using D-optimal designs to examine how many satellite tags are required to estimate dispersal rates between metapopulations. Examining these questions \emph{a priori} and as part of the design of field programs, and even the vetting of proposals, ought to become standard practice. However, as yet, the minutiae of such approaches are not being considered. We feel this is an important missing link in the dialogue between statisticians and ecological researchers. At present, analysis of animal movement data is often reactive or opportunistic, and the statistics are seen as a secondary step, only to be engaged in once data is streaming in from tagged individuals in the field. A recent paper \citep{mcgowan2016integrating} has considered how conservation efforts can usefully use animal tracking data. Such assessments are not new in other areas of applied ecology---evaluation of tagging studies has been frequently employed in fisheries management \citep[e.g. see][]{sippel2015using, eveson2012using}. We hope this paper is indicative of a trend toward more quantitative assessments of the design of tagging studies.   

\subsection{Retrograde steps in movement analysis}
\label{levy}

Animal movement patterns continue to be analysed by an array of new statistical techniques that seek to classify behavioural states \citep[e.g.][]{madon2014deciphering, sur2014change, zhang2015extending}. Those that we have discussed here are merely the tip of the iceberg, relative to the number available in the ecological literature. The approaches considered here were chosen because we feel that they are currently the best statistical approaches for analysis of behaviour in relatively high-accuracy individual movement tracks. We do not have space to consider the larger literature here, but there are some general points to be made about model complexity. In this section we consider the general pitfalls of using what many consider to be too simple a model. 

An example of a relatively simple statistical approach that has gained high prominence (see, e.g., \citealp{sims2008scaling, humphries2010environmental, de2011levy}) is the L\'evy-flight (or L\'evy-walk) hypothesis. This was first proposed as ubiquitous model of random searching behaviour by \citet{viswanathan1999optimizing}.  Under this model a simple power function is used to model the distribution of step lengths. Several papers have argued for the ubiquity of the model and claimed that animal search strategies that employ stochastic movements in accordance with the model will be optimal (i.e.\ lead to the best foraging success in the long run relative to a Brownian motion model). But the approach has also courted controversy \citep{Edwards2007, Edwards2011, Pyke2015}. \citet{Pyke2015} criticizes the approach on theoretical and largely non-statistical grounds, and states that the controversies regarding statistical methods for L\'evy flights, as discussed e.g.\ in \citet{Edwards2007}, are largely a ``red-herring''. This may be true in this particular case. Nonetheless, if statistical inference is used to guide scientific inference, then it is critical to align the statistical models with the biological inference so that the two are commensurate and that biological conclusions are well supported by empirical evidence \citep{edwards2012incorrect}. We believe there are two critiques of L\'evy-like approaches that are not generally appreciated by ecologists. 
The temporal dependence in movement data is a crucial factor that must be accounted for in inference and model selection. Since movement data is often fundamentally auto-correlated, the samples from any distribution of step lengths are not independent. This is ignored at the researcher's peril when attempting to discriminate between candidate models/hypotheses. Spurious tests of significance are highly likely and the size of animal movement data sets increases the power of any significance test to discriminate between what could be biologically irrelevant differences. This criticism applies to both sides of the debate around L\'evy-like models. \citet{Edwards2007}, in an article critical of much of the foregoing estimation of L\'evy models, details likelihood based approaches but largely ignores the crucial issue of auto-correlation.  This is why such importance should be placed on time series approaches. The methods considered which do account for temporal dependence are still merely caricatures of the real processes that influence how animals make decisions about movement. Nonetheless they constitute progress towards capturing reality, and are often sufficient for capturing the broad features in the data via a time-dependent (e.g.\ using Markov assumptions) likelihood function.  

In that sense, we, along with other authors \citep{Pyke2015}, question whether the strong claims about animal movement (such as ubiquity and optimality) have been strongly tested by the use of simple models such as those based on power laws.  The set of candidate models in such papers is often pitifully small (e.g.\ L\'evy vs.\ Brownian models), and little is to be gained by comparing the merits of two obviously over-simplified models. In addition, simple models are highly limited in their ability to encode the results of previous studies. In comparison, and simply as an example, the HMM framework we considered here can incorporate factors such as energetic reserves altering behavioural choices  \citep{Zucchini2009}, environmental drivers \citep{Patterson2009} and individual variation \citep{Langrock2012}. 


\subsection{Dependence between individuals}
The vast majority of statistical analysis of movement focuses on one individual at a time. Historically, this made sense give the sparsity of data, but modern global positioning system (GPS) technology means that it is increasingly common to have multiple individuals that are simultaneously tracked and potentially interdependent, either because they interact directly or because they respond to the same events or variations in their environment. 
Statistical treatment of this case lags behind the data collection, notwithstanding the early mention by \citet{Dunn1977} and recent papers using approaches outside the set of statistical techniques considered here \citep{scharf2015dynamic,russell2016dynamic}. Within the HMM context \citet{Langrock2014} described a  discrete-time model with state switching, building on the ideas in Section \ref{HMMs} to describe intermittent dependence between animals in a parsimonious way. \citet{NiuGroup} developed this approach in continuous time using the diffusion approach of Section \ref{DM}, jointly representing the locations of interacting individuals as the state of a single high-dimensional process. Appropriate allowance for dependence is crucial for statistically-valid exploitation of data currently being generated.

%
%

\subsection{Interpretation of model states in HMMs and SSMs}
\label{states}
In the literature on animal movement modelling using state-switching models, whether HMMs, SSMs or diffusions, the states of the Markov chain are often interpreted as behavioural states of the animals considered. While the general sentiment of relating the models' states to the animals' underlying motivation clearly makes sense, there is, in our view, a tendency to over-interpret these types of models. The probabilistic features within states are data driven in the sense that when fitting the model to the data there is no mechanism that would guarantee that the patterns that will be picked up by the model are in any way biologically meaningful with regard to behavioural states. The model simply picks up the strongest patterns in the data. For example, there could be three biologically meaningful behavioural states, say resting, feeding and travelling, but the fitted three-state model is such that resting and feeding are lumped together into one model state (due to the step lengths and turning angles being of similar magnitude) while short-distance and long-distance travelling activities are differentiated by the remaining two states (possibly due to insufficient flexibility of the state-dependent distributions considered). Perhaps even more importantly, the temporal resolution of the observations will often not allow for any direct interpretation of the states (e.g.\ if time intervals between successive observations are such that animals will often exhibit several different behaviours within each interval).

\citet{BlackwellExact} give examples of various kinds. In their spatially-heterogeneous model of fisher movement, the  model states are constrained to have a one-to-one correspondence with habitat types, and are shown to have a much improved fit compared with a spatially-homogeneous model, strongly suggesting that the states represent some biologically meaningful aspect of movement behaviour in response to environment. Their analysis of wild boar movement has a small supervised element, partly to tackle statistical issues with label-switching; 
this also has the effect of ensuring that particular states correspond to some features of the data that are both obvious and biologically interpretable, namely clusters of locations during periods of inactivity. 
The same analysis also illustrates the caveats above; some of the states, originally envisaged as ``foraging'' and ``travelling'' states, clearly improve the fit of the model, but their pattern of occurrence suggests that they do not have such a direct interpretation, but rather capture less interpretable heterogeneity in movement over time. 



\subsection{Final remarks}

Although considerable progress has been made over the last decade, the development of statistical tools for modelling animal movement data is only beginning to catch up with the explosion in the volume of corresponding data and associated modelling challenges. There has been, and still is, a huge demand for statistical expertise. Crucially, we believe that the end-users (i.e.\ ecologists) mostly do not need more sophisticated but case-specific and technically intractable models, but instead need intuitive and practical tools which they can implement and handle themselves. The main challenge clearly lies in identifying the right balance between overly complex yet inaccessible and accessible yet overly simplistic modelling approaches. Progress toward this end will require close collaboration between statisticians and movement ecologists. Such a process ought to inform all parts of movement ecology, from the design of tags, sensors and instruments, to study design and ultimate analysis.

\subsection*{Acknowledgments}
We thank Geoff Hosack and two anonymous referees for their useful feedback on this manuscript.

\renewcommand\refname{References}
\makeatletter
\renewcommand\@biblabel[1]{}

\bibliography{ReferencesPartonLangrock,extras}
\bibliographystyle
{abbrvnat}

\end{spacing}

\end{document}